%% file: DominguezGuzman.tex
\documentclass[a4paper,fleqn,usenatbib]{mnras}

\usepackage{newtxtext,newtxmath}

\usepackage[T1]{fontenc}
\usepackage{ae,aecompl}

\usepackage{graphicx}	
\usepackage{amsmath}	
\usepackage{pdflscape}
\usepackage{rotating}  

\title[The chemical homogeneity of H\,{\normalsize \textit{II}} regions in the MCs]{The homogeneity of chemical abundances in H\,{\Large \textbf{II}} regions of the Magellanic Clouds}

\author[Dom\'inguez-Guzm\'an et al.]{G.\ Dom\'inguez-Guzm\'an,$^{1}$\thanks{E-mail: gisedogu@inaoep.mx} M.\ Rodr\'iguez,$^{1}$ J.\ Garc\'ia-Rojas,$^{2,3}$ C.\ Esteban,$^{2,3}$ 
\newauthor and L.\ Toribio San Cipriano$^4$ \\
$^{1}$Instituto Nacional de Astrof\'isica, \'Optica y Electr\'onica, Luis Enrique Erro 1, Tonantzintla 72840, Puebla, Mexico \\
$^{2}$Instituto de Astrof\'isica de Canarias, E-38200, La Laguna, Tenerife, Spain \\
$^{3}$Departamento de Astrof\'isica, Universidad de La Laguna, E-38206, La Laguna, Tenerife, Spain \\
$^{4}$Centro de Investigaciones Energ\'{e}ticas, Medioambientales y Tecnol\'{o}gicas (CIEMAT), Avda. Complutense 40, E-28040, Madrid, Spain \\
}

\date{Accepted XXX. Received YYY; in original form ZZZ}

\pubyear{2022}

\begin{document}
\label{firstpage}
\pagerange{\pageref{firstpage}--\pageref{lastpage}}
\maketitle

\begin{abstract}
We use very deep spectra obtained with the Ultraviolet-Visual Echelle Spectrograph at the Very Large Telescope to derive physical conditions and chemical abundances of four \ion{H}{ii} regions of the Large Magellanic Cloud (LMC) and four \ion{H}{ii} regions of the Small Magellanic Cloud (SMC). The observations cover the spectral range 3100--10400~\AA\ with a spectral resolution of $\Delta\lambda\ge\lambda/11600$, and we measure 95--225 emission lines in each object. We derive ionic and total abundances of O, N, S, Ne, Ar, Cl, and Fe using collisionally excited lines. We find average values of $12+\log(\mbox{O/H})=8.37$ in the LMC and $8.01$ in the SMC, with standard deviations of $\sigma=0.03$ and 0.02~dex, respectively. The S/O, Ne/O, Ar/O, and Cl/O abundance ratios are very similar in both clouds, with $\sigma=0.02$--0.03~dex, which indicates that the chemical elements are well mixed in the interstellar medium of each galaxy. The LMC is enhanced in N/O by $\sim0.20$~dex with respect to the SMC, and the dispersions in N/O, $\sigma=0.05$~dex in each cloud, are larger than those found for the other elements. The derived standard deviations would be much larger for all the abundance ratios, up to 0.20~dex for N/O, if previous spectra of these objects were used to perform the analysis. Finally, we find a wide range of iron depletions in both clouds, with more than 90 per cent of the iron atoms deposited onto dust grains in most objects.
\end{abstract}




\section{Introduction}\label{sec1}
The Magellanic Clouds (MCs) are two irregular galaxies, the Large Magellanic Cloud (LMC) and the Small Magellanic Cloud (SMC), that orbit the Milky Way. Because of their proximity and low-metallicity, they provide an excellent opportunity to explore in detail the chemical composition of \ion{H}{ii} regions in dwarf galaxies and at different metallicities.

However, it is difficult to obtain reliable optical spectra of MC \ion{H}{ii} regions since these galaxies are generally observed at high airmasses, making the observations very sensitive to the effects of atmospheric differential refraction \citep{Fil82}. Besides, most of the available spectra are not very deep. Of all the available spectra of MC \ion{H}{ii} regions \citep{PT74,D75,PEFW78,STHM86,Kurt99,GGC00,TBLDS03,NRMCV03,PA03}, only the spectrum of 30 Doradus presented by \cite{PA03} can be considered of a particularly high quality since it is a deep spectrum with high wavelength coverage and relatively high spectral resolution that was obtained using an atmospheric dispersion corrector. In this work, we analyse eight new spectra of MC \ion{H}{ii} regions that have a comparable quality.

Dwarf irregular galaxies are generally considered to be chemically homogeneous \citep{KS97,Lee06,Crox09}, although some studies find evidence of metallicity gradients or inhomogeneities \citep{Pilyu15,Anni2017,James16,James20}. \citet{PEFW78} determined oxygen abundances in several MC \ion{H}{ii} regions and found a flat gradient for oxygen in the SMC, with abundance variations compatible with the observational errors, but also found some marginal evidence for a gradient in the LMC. More recently, in a pre-analysis of the spectra presented here, \citet{TSC17} study the spatial distribution of oxygen as a function of the galactocentric distance in the MCs. They use collisionally excited and recombination lines of oxygen and carbon and find that the radial abundance gradient is practically flat for both galaxies, and that the abundance variations are small. However, \citet{Roman21} use the interstellar depletions derived for different elements towards many lines of sight in the LMC to infer variations in metallicity of up to 0.8~dex across the LMC. 

Here we revisit this issue. We present the full observed spectra of the LMC \ion{H}{ii} regions IC~2111, N11B, N44C, and NGC~1714, and the SMC \ion{H}{ii} regions N66A, N81, N88A, and N90, all of them previously pre-analysed by \citet{TSC17}. We use these spectra to determine the physical conditions and chemical abundances of He, O, N, S, Ne, Ar, Cl, and Fe. Our aim is to determine the best estimates of these parameters and to constrain the variations in chemical abundances across the MCs. We also provide a comparison between our results and those that would be obtained from previous spectra of lower quality.

\section{Observations and data reduction}\label{sec2}

\begin{figure}
	\center
	\includegraphics[width=0.44\textwidth, trim=10 7 5 5, clip=yes]{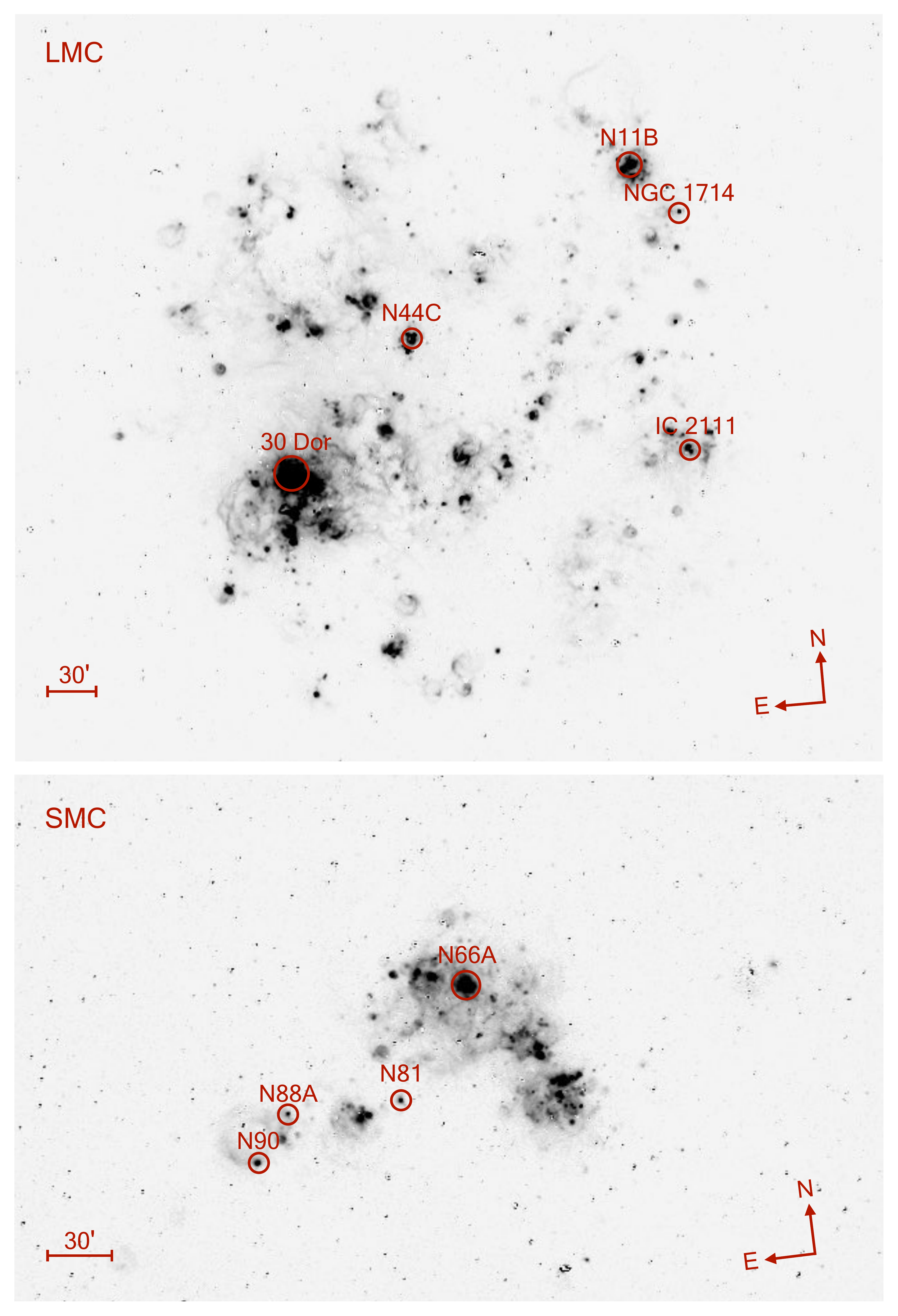}
	\caption{H$\alpha$ images from the Southern H$\alpha$ Sky Survey Atlas \citep[SHASSA;][]{Gaus01} for the LMC (top panel) and the SMC (bottom panel). The positions of the \ion{H}{ii} regions studied in this work are indicated with circles.}
	\label{fig1}
\end{figure}

We obtained echelle spectra of eight \ion{H}{ii} regions in the MCs, four in the SMC and four in the LMC, on 2003 March 30--31 (IC~2111 and NGC~1714) and 2014 November 3--4 (rest of the objects) with the Ultraviolet-Visual Echelle Spectrograph \citep[UVES;][]{uvespaper} at the Very Large Telescope (VLT; Kueyen Telescope) in Cerro Paranal Observatory, Chile. Fig.~\ref{fig1} shows the distribution of the observed \ion{H}{ii} regions in each galaxy. We also show 30~Dor in this figure because it is included in the analysis described below.

The observations were carried out with the standard settings of the spectrograph, using the blue and red arms for each dichroic beam splitter. The complete spectra are divided in four sections: the blue ranges 3100--3880~\AA\ (B1) and 3760--4986~\AA\ (B2), and the red ranges 4785--6828~\AA\ (R1) and 6700--10420~\AA\ (R2). The total spectral range covered goes from $3100$~\AA\ to $10420$~\AA, with a spectral resolution of $\Delta\lambda\sim\lambda/20000$ for NGC~1714 and $\Delta\lambda\sim\lambda/11600$ for the rest of the objects. Two spectral intervals were not observed, 5783--5820~\AA\ and 8540--8650~\AA, due to the physical separation between the two CCDs used in the red arm. In addition, five smaller regions were not observed, 9608--9620~\AA, 9761--9775~\AA, 9918--9935~\AA, 10080--10100~\AA\ and 10249--10272~\AA, because the last two orders do not fit within the CCD.

\begin{figure}
	\center
	\includegraphics[width=0.42\textwidth, trim=10 7 5 5, clip=yes]{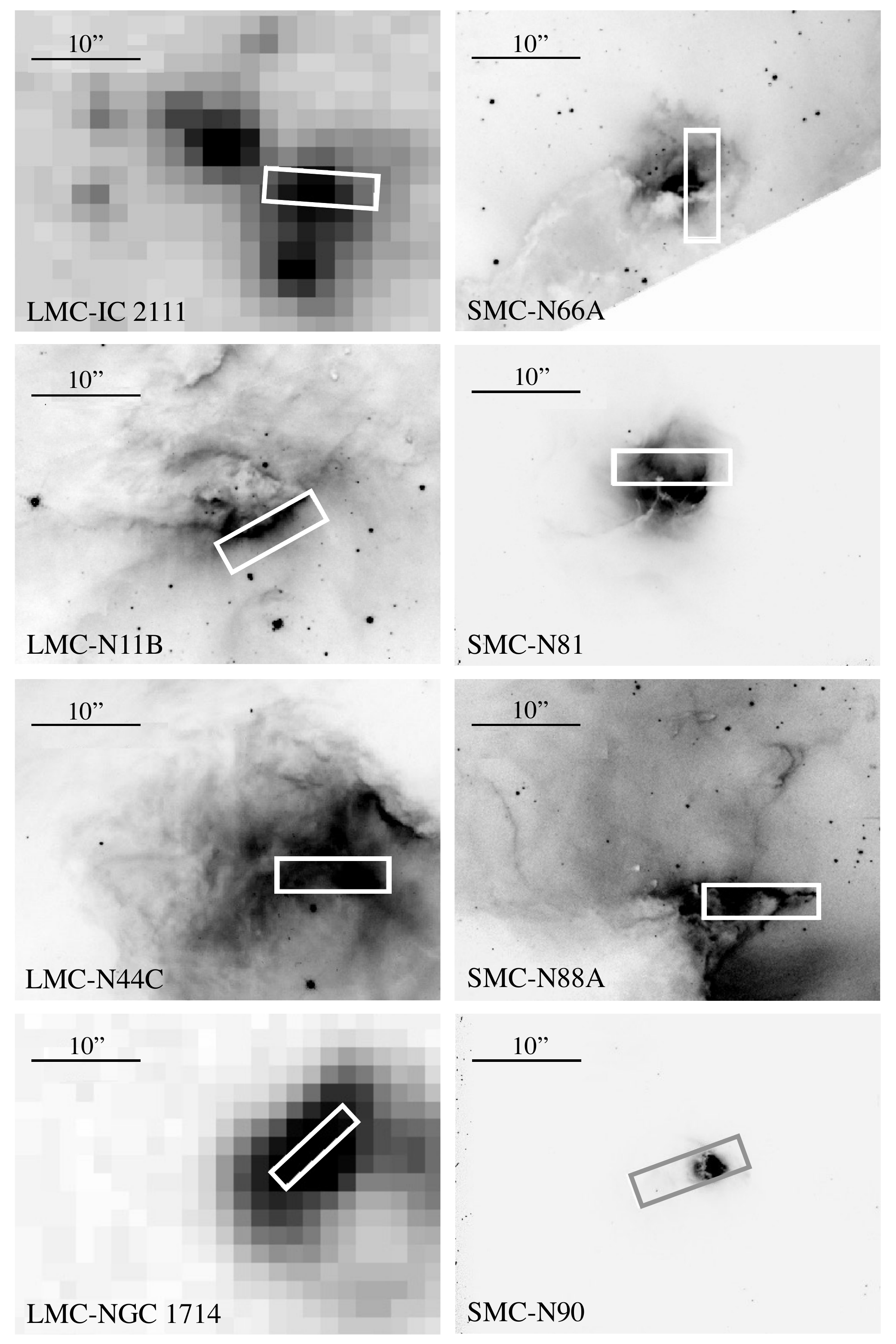}
	\caption{H$\alpha$ images for N11B, N44C, N66A, N81, N88A, and N90 obtained from HST/HDA. The R band images for NGC~1714 and IC~2111 were taken from `Aladin sky atlas' \citep{ASA00}. The rectangles indicate the positions and sizes of the slit. North is to the top and east to the left.}
	\label{fig2}
\end{figure}

The slit width was set to 2 arcsec for NGC~1714 and 3 arcsec for the rest of the objects, and the slit length is equal to 9.5 arcsec in the blue arm and 11.5 arcsec in the red arm. The slit was set at different position angles (PAs) trying to cover the brightest areas of the objects. Fig.~\ref{fig2} shows the positions and sizes of the slit overplotted in R band images from `Aladin sky atlas' \citep{ASA00} for NGC~1714 and IC~2111, and in H$\alpha$ images from HST/HDA for the rest of the objects. The atmospheric dispersion corrector was used to keep the same observed region within the slit at different wavelengths (the MCs are observed at relatively high airmasses, between 1.4 and 2.0). The seeing during the observations was better than $\sim1$ arcsec.

The technical details of the observations, such as coordinates, exposure times, position angles, extracted areas, and airmasses, are provided in Table \ref{table1}, where $z$ is the zenith angle.

\begin{table*}
 \caption{Journal of observations.} 
  \label{table1}
 \begin{tabular}{l l c c c c c c c}
  \hline
  & Object & R.A. (J2000) & Dec. (J2000)& \multicolumn{2}{c}{Exposure time (s)} & PA & Extracted area & Airmass\\
  Galaxy & name & (hh:mm:ss) & ($\degr~\arcmin~\arcsec$) & (B1, R1) & (B2, R2) & ($\degr$)& (arcsec$^2$) & $\sec(z)$ \\
  \hline 
  LMC & IC~2111$^\mathrm{a}$ & 04:51:52.1 & -69:23:32.0 & $1\times60,~3\times240$ & $1\times60,~3\times700$  & 90 & $3.0\times9.5$ &1.8 - 2.0\\ 
  LMC & N11B$^\mathrm{b}$ & 04:56:46.9 & -66:24:37.9 & $3\times30,~3\times900$ & $3\times30,~4\times2000$ & 120 & $3.0\times9.5$ & $\sim1.4$\\ 
  LMC & N44C$^\mathrm{b}$ & 05:22:13.6 & -67:58:34.2 & $3\times30,~3\times300$ & $3\times30,~3\times1200$ &  90  & $3.0\times9.5$ &  $\sim1.4$\\ 
  LMC & NGC~1714$^\mathrm{a}$ & 04:52:08.8 & -66:55:24.0 & $3\times300$ & $3\times900$  & 315 & $2.0\times9.5$ &1.6 - 1.9\\ 
  SMC & N66A$^\mathrm{b}$ & 00:59:14.3 & -72:11:02.8 & $3\times30,~3\times600$ & $3\times30,~3\times1500$ &  0    & $3.0\times9.5$ & 1.5 - 1.6\\ 
  SMC & N81$^\mathrm{b}$ & 01:09:12.8 & -73:11:36.9 & $3\times30,~3\times600$ & $3\times30,~3\times1800$ &  90  & $3.0\times9.5$ & 1.5 - 1.7\\  
  SMC & N88A$^\mathrm{b}$ & 01:24:08.3 & -73:09:04.6 & $3\times30,~3\times800$ & $3\times30,~3\times1200$ &  110 & $3.0\times5.3$ & 1.5 - 1.7\\  
  SMC & N90$^\mathrm{b}$ & 01:29:36.1 & -73:33:51.9 & $3\times30,~3\times200$ & $3\times30,~2\times1200,1800$ &  90  & $3.0\times9.5$ & 1.5 - 1.6\\  
  \hline 
  \multicolumn{9}{l}{$^\mathrm{a}$Observation dates: 2003 March 30 and 31.}\\
  \multicolumn{9}{l}{$^\mathrm{b}$Observation dates: 2013 November 3 and 4.}
 \end{tabular}
\end{table*}

Data reduction was performed using the public UVES pipeline under the {\sc gasgano} graphic user interface \citep{gasgano}, which includes the tools for the standard procedures of bias subtraction, flat fielding and wavelength calibration. The final results of the pipeline are 2D wavelength-calibrated spectra. For flux calibration we use the available tasks in the {\sc iraf}\footnote{{\sc iraf} is distributed by the National Optical Astronomy Observatories, which are operated by the Association of Universities for Research in Astronomy, Inc., under cooperative agreement with the National Science Foundation.} software package using the standard stars HR718, HR3454 and HR9087 \citep{HWSGHP92,Ham94}, which can be used in the wavelength range 3300--10500~\AA. The error associated to the flux calibration is calculated using the standard deviation of the fitted sensitivity function, which is equal to 4 per cent for NGC~1714 and IC~2111, and 1 per cent and 2.5 per cent for the ranges B1-R1 and B2-R2, respectively, in the other objects.

The [\ion{O}{iii}] $\lambda\lambda4949$, 5007 lines are saturated in the long-exposure spectra of N81, and the [\ion{O}{iii}] $\lambda\lambda4949$, 5007 and H$\alpha$ lines are saturated in the long-exposure spectra of N88A. We have measured these lines in the short-exposure spectra, with the measurements normalized to those obtained in the long-exposure spectra using the lines that can be measured well in both cases.

\begin{figure*}
	\includegraphics[width=15cm]{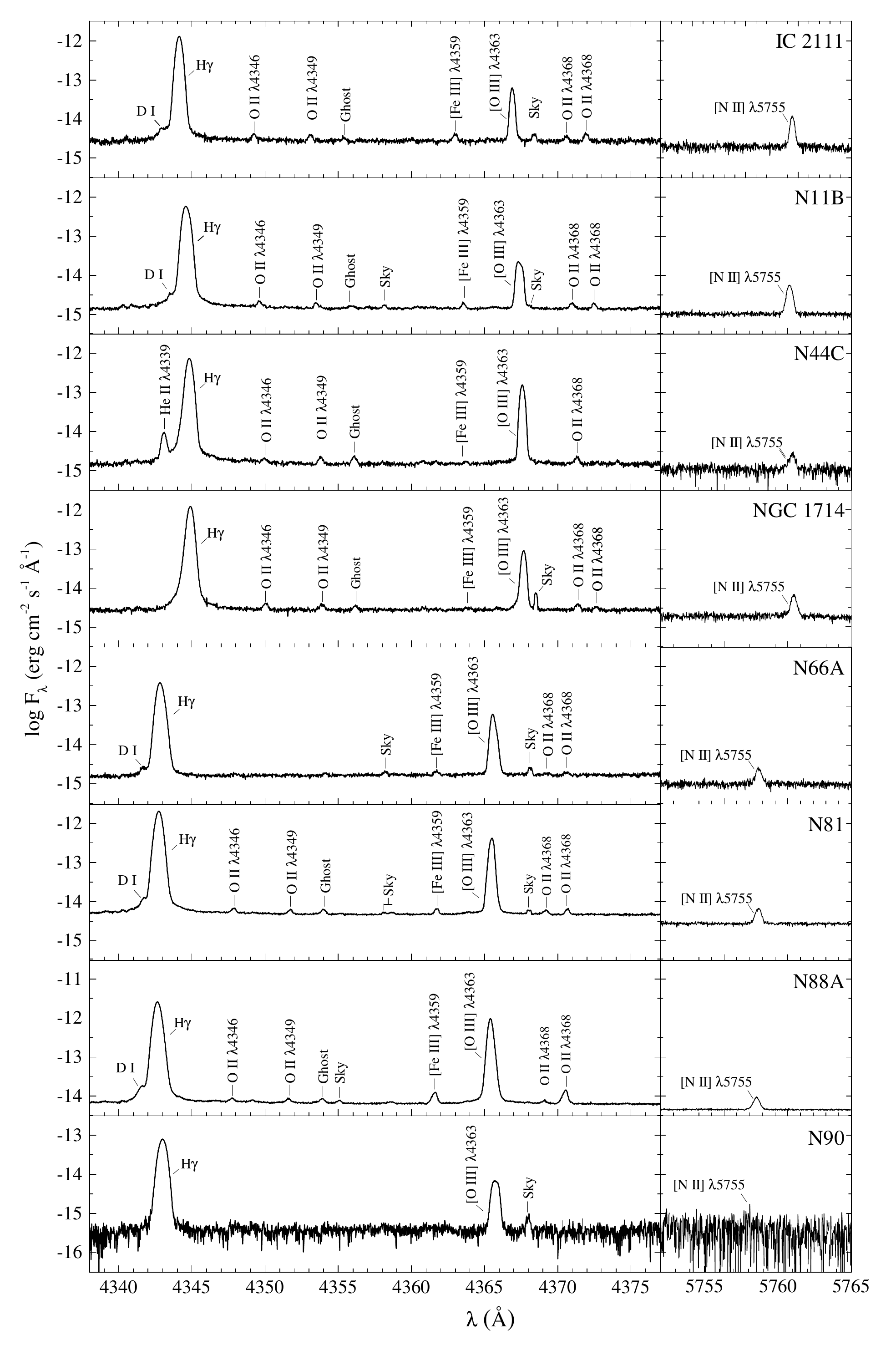}
	\caption{Parts of the observed spectra for each object, illustrating the (mostly) high signal-to-noise detections of [\ion{O}{iii}] $\lambda4363$ and [\ion{N}{ii}] $\lambda5755$, the lines that we use to determine the electron temperature. The wavelengths are not corrected to their rest values.}
	\label{fig3}
\end{figure*}

Fig.~\ref{fig3} shows parts of the flux calibrated spectrum for each object, where the auroral [\ion{O}{iii}] $\lambda4363$ and [\ion{N}{ii}] $\lambda5755$ lines, which are important to determine the electron temperature, can be seen to have very high signal-to-noise ratios in most cases. The main exception is N90, where the exposure times were not long enough to detect [\ion{N}{ii}] $\lambda5755$.

\section{Flux measurements and reddening correction}\label{sec3}

The line intensities have been measured by integrating the flux above the continuum defined by two points on each side of the emission lines using the task {\sc splot} of {\sc iraf}, except for the blended emission lines, where we fitted Gaussian profiles. We are able to measure between 92 and 225 emission lines in each object. The uncertainties associated to these measurements are estimated using the equation given by \citet{TMLS99}, which adds quadratically the uncertainties introduced by the measurement of both the line and the continuum.

The reddening coefficient, $c(\mbox{H}\beta)$, is determined by comparing the intensities of several Balmer and Paschen lines relative to H$\beta$ with their case~B values \citep{SH95}. We exclude those lines that are blended with other lines or affected by telluric emission, and use lines whose upper levels have principal quantum numbers $n\le7$, since for $n>7$ the lines depart from their expected case~B values, as illustrated below. This behaviour was previously found in the Orion Nebula by \cite{MD09}, who argue that it could arise from collisions that change the quantum number $l$ by more than $\pm1$ or from pumping of the \ion{H}{i} lines by absorption of the stellar continuum.

\begin{figure}
	\center
	\includegraphics[width=0.42\textwidth, trim=10 8 5 5, clip=yes]{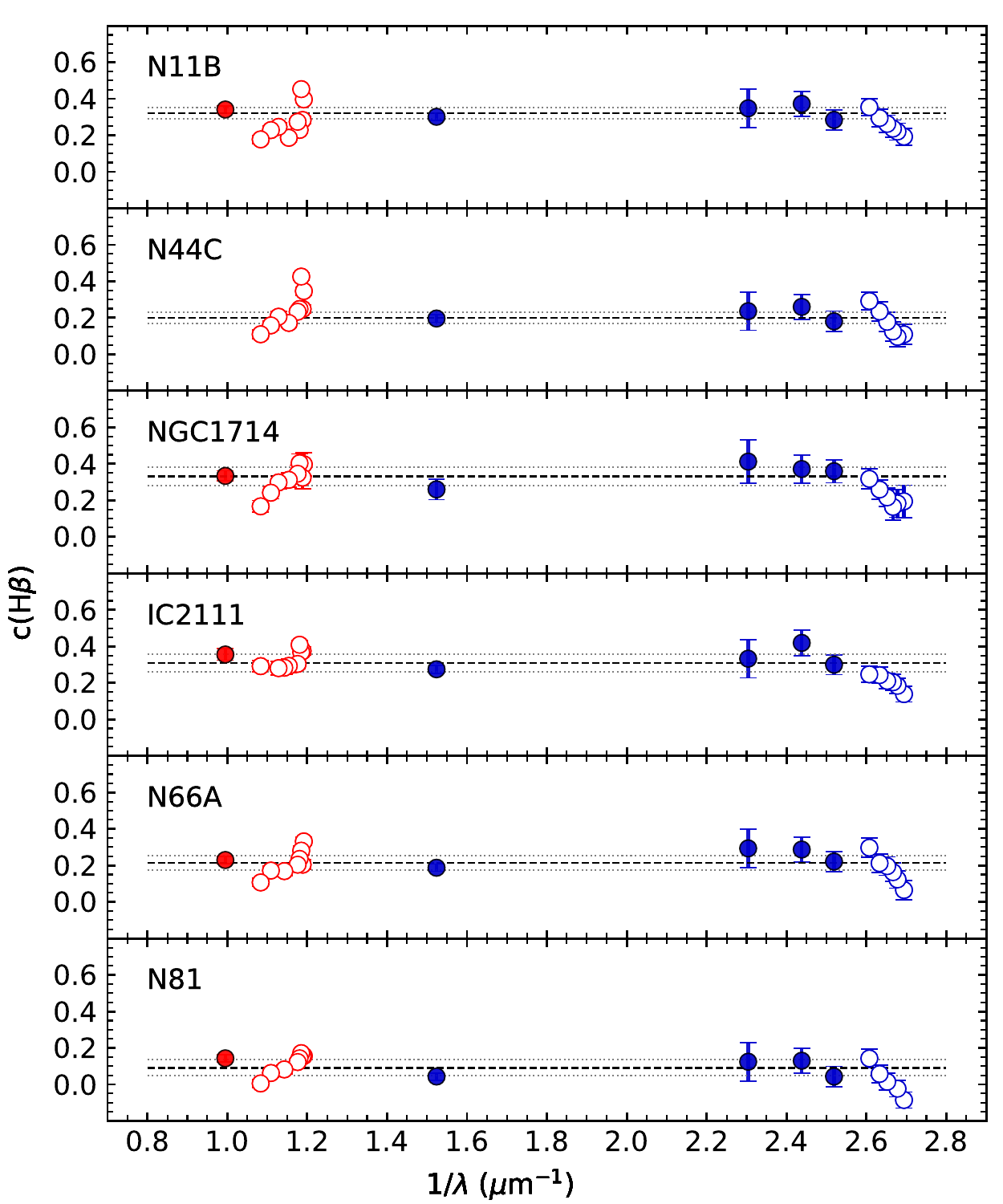}
	\caption{Reddening coefficients, $c(\mbox{H}\beta)$, implied by the intensities of different \ion{H}{i} lines relative to H$\beta$ as a function of the inverse wavelength of these lines in $\mu$m. Blue/dark circles show the results obtained with the Balmer lines and red/light circles those implied by the Paschen lines. Filled circles indicate the lines that we use to estimate the final value of $c(\mbox{H}\beta)$ using a weighted mean (long dashed line) and its standard deviation (small dashed lines).}
	\label{fig4}
\end{figure}

We use the reddening law of \citet{H83}, which has a ratio of total to selective extinction $R_{\mbox{v}}=3.1$ and is commonly used for the MCs, for all objects excepting N88A and N90. Fig.~\ref{fig4} shows the values of $c(\mbox{H}\beta)$ implied by the different \ion{H}{i} line ratios as a function of the inverse wavelength for the six regions where the reddening law of \citet{H83} leads to consistent results. The values of $c(\mbox{H}\beta)$ implied by the intensities of the Balmer and Paschen lines relative to H$\beta$ are shown with blue and red circles (dark and light grey circles in the printed version), respectively. The results for lines whose upper levels have $n\le7$, which are shown with filled circles in Fig.~\ref{fig4}, are used to estimate the weighted means (indicated by the long dashed lines) and the standard deviations (small dashed lines). The results for lines whose upper levels have $n>7$, which are represented with open circles, can be seen to deviate from their expected values.

\begin{figure}
	\center
	\includegraphics[width=0.42\textwidth, trim=10 8 5 5, clip=yes]{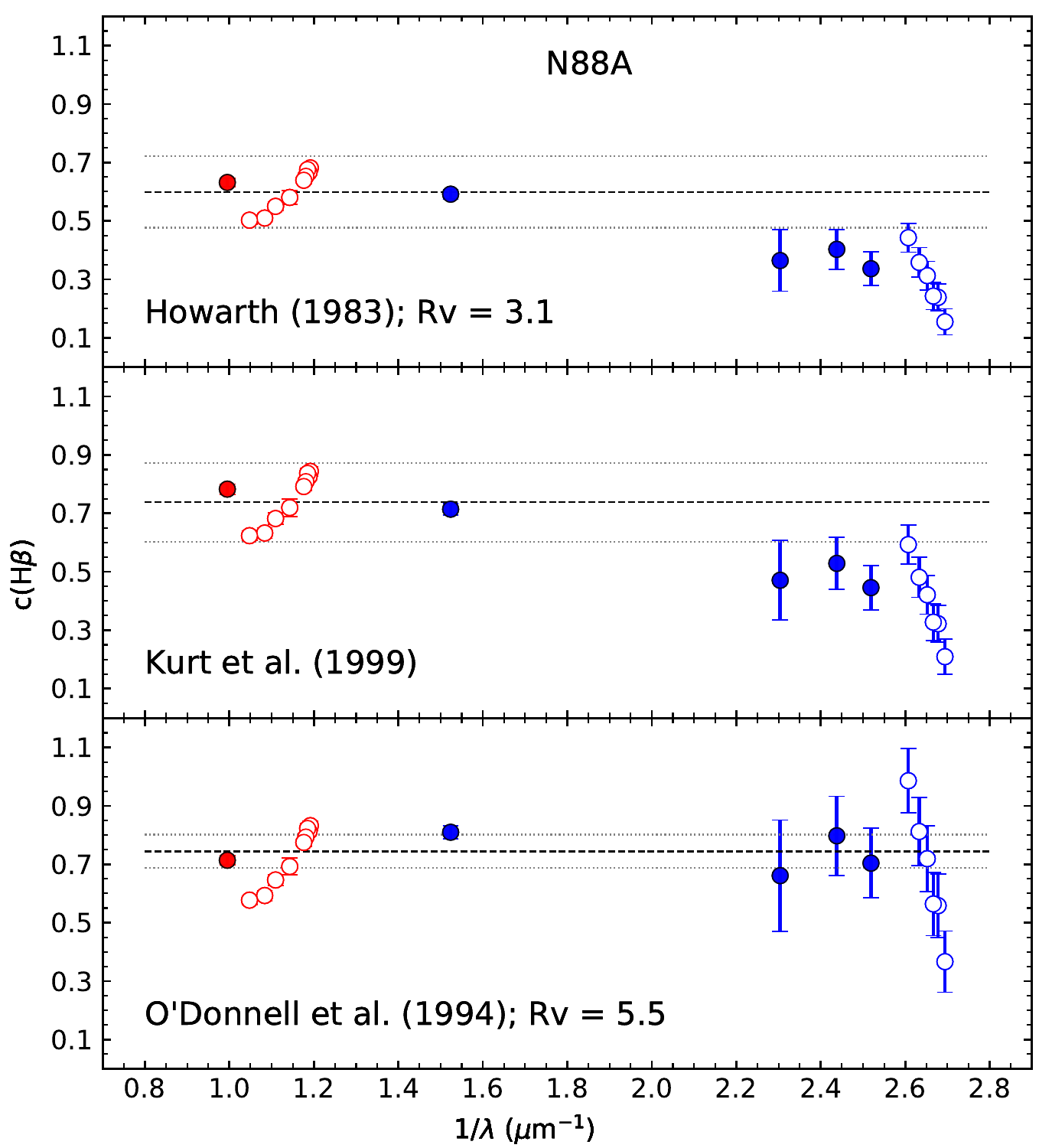}
	\caption{Reddening coefficient $c(\mbox{H}\beta)$ as a function of the inverse wavelength in $\mu$m for three extinction laws: \citet{H83} (top panel), \citet{Kurt99} (middle panel) and \citet{OD94} with $R_{\mbox{v}}=5.5$ (bottom panel) for N88A. Blue/dark circles show the results obtained with the Balmer lines and red/light circles those implied by the Paschen lines. The filled circles are those values we use to estimate the weighted mean (long dashed line) and the standard deviation (small dashed lines).}
	\label{fig5}
\end{figure}

\begin{figure}
	\center
	\includegraphics[width=0.42\textwidth, trim=10 8 5 5, clip=yes]{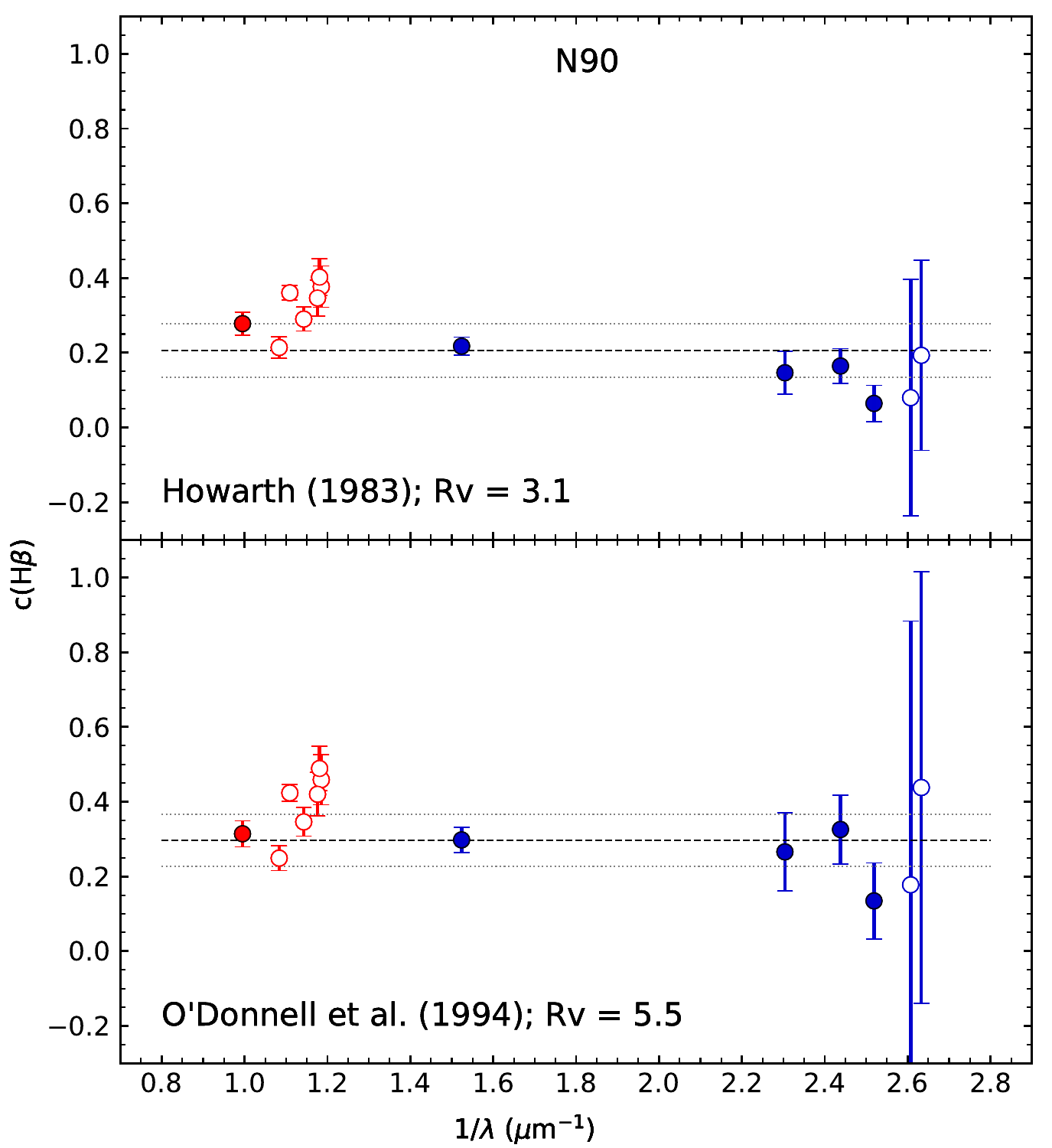}
	\caption{Reddening coefficient $c(\mbox{H}\beta)$ as a function of the inverse wavelength in $\mu$m for the extinction laws of \citet{H83} (top panel) and \citet{OD94} with $R_{\mbox{v}}=5.5$ (bottom panel) for N90. Blue/dark circles show the results obtained with the Balmer lines and red/light circles those implied by the Paschen lines. The filled circles are those values we use to estimate the weighted mean (long dashed line) and the standard deviation (small dashed lines).}
	\label{fig6}
\end{figure}

The upper panels of Figs.~\ref{fig5} and \ref{fig6} show the corresponding results for N88A and N90: the values of $c(\mbox{H}\beta)$ implied by the different \ion{H}{i} line ratios as a function of the inverse wavelength when the extinction law of \citet{H83} is used. There is a trend in both cases of $c(\mbox{H}\beta)$ with the inverse wavelength, with the Paschen lines providing higher values of $c(\mbox{H}\beta)$ than the Balmer lines. We tested other extinction laws in order to see which ones worked better for these objects. We find that the law of \citet{OD94} for $R_{\mbox{v}}=5.5$ provides reasonable fits for both objects. The results implied by this law are plotted in the bottom panels of Figs.~\ref{fig5} and \ref{fig6}. We also show the results implied by the law of \citet{Kurt99} for N88A (middle panel of Fig.~\ref{fig5}), since this extinction law was derived specifically for this object, but it can be seen in this figure that the law of \citet{OD94} works better.

The effect of changing the extinction law and the corresponding value of $c(\mbox{H}\beta)$ is significant for several of the quantities that we will be calculating. In the case of N88A, where the effects are larger because of the larger value of $c(\mbox{H}\beta)$, if the law of \citet{H83} were used instead of the one by \citet{OD94}, O/H would decrease by 0.03~dex and the abundances of the other elements relative to oxygen would increase by up to 0.12~dex, being N/O the most affected abundance ratio. Part of these changes arise from changes in the electron temperatures used to calculate the abundances, $T_{\text{e}}$([\ion{N}{ii}]) and $T_{\text{e}}$([\ion{O}{iii}]), which would increase with the \citet{H83} law by 80 and 410~K, respectively. The other estimates of the electron temperature that can be calculated with our spectra for this object, $T_{\text{e}}$([\ion{S}{ii}]), $T_{\text{e}}$([\ion{O}{ii}]), $T_{\text{e}}$([\ion{S}{iii}]), and $T_{\text{e}}$([\ion{Ar}{iii}]), would decrease by 3700, 2900, 900, and 1400~K, respectively. Since the results in Figs.~\ref{fig5} and \ref{fig6} show that the extinction law of \citet{OD94} is working much better both in N88A and in N90, the effect of changing the extinction law is not included in our estimates of the uncertainties in the final line intensity ratios.

The final uncertainty associated to the measurement of each emission line is estimated by adding quadratically the uncertainty in the flux calibration, the uncertainty in the measured intensity, and the uncertainty in $c(\mbox{H}\beta)$. The emission line intensities, which are not corrected for telluric absorption, are listed in Appendix~\ref{appendixA}, Tables~\ref{tableLMC1} to \ref{tableSMC2}. These tables provide the values of the emission line intensities uncorrected, $F(\lambda)$, and corrected, $I(\lambda)$, for reddening for each emission line. We include the laboratory wavelengths $\lambda_0$, the ions, the multiplet numbers (ID), the extinction law $f(\lambda)$, the observed wavelengths in the heliocentric framework $\lambda$, the observed fluxes in H$\beta$, and the reddening coefficients $c(\mbox{H}\beta)$.  

A preliminary analysis of these spectra, centered on the determination of the oxygen and carbon abundances from \ion{O}{ii} and \ion{C}{ii} recombination lines was presented in \citet{TSC17}. The intensities of several emission lines were reported by \citeauthor{TSC17}, and they are slightly different from those presented here because we have made some improvements in the reduction procedure and in the reddening correction. Some of the atomic data and the procedure used to calculate the physical conditions and chemical abundances are also different here. However, the differences between the results of \citeauthor{TSC17} and those presented here for the total abundances of oxygen derived with collisionally excited lines are practically negligible, since they are equal or lower than $\sim0.01$~dex for all objects excepting N88A, where the difference reaches 0.08~dex.

\section[]{Previous spectra of the H\,{\sevensize II} regions in our sample}\label{sec4}

In order to explore the impact that high-quality spectra have in the determinations of chemical abundances, we have compiled from the literature 19 previous spectra of the \ion{H}{ii} regions in our sample \citep{PT74,D75,PEFW78,STHM86,Kurt99,GGC00,TBLDS03,NRMCV03,PA03}. Of these spectra, the one of 30~Dor presented by \citet{PA03} was also observed with UVES/VLT, has similar quality to those presented here, and is included as an additional object to our main sample of VLT spectra. Each spectrum has at least one density and temperature diagnostic and the emission lines needed to determine the abundances of several elements. We use the reddening-corrected line intensities relative to H$\beta$ listed in each work. All the spectra, the ones presented here and the ones obtained from the literature, are analysed in an homogeneous way following the procedure described below.

\section{Analysis}\label{sec5}

The calculations of physical conditions and ionic abundances are done with {\sc pyneb} v1.1.15 \citep{LMS15}, a Python based code that is used for the analysis of emission lines. The atomic data set used for the analysis of the collisionally excited lines is shown in Table~\ref{table2}. These atomic data do not include any of the problematic datasets identified by \citet{JDR17,JDR21}. For the analysis of the \ion{H}{i}, \ion{He}{i}, and \ion{He}{ii} recombination lines, we use the recombination coefficients of \citet{SH95} and \citet{PFSD12}.

The uncertainties are obtained through Monte Carlo simulations. For each line, we generate 10000 random values from a Gaussian distribution centered in the observed line intensity and with a standard deviation equal to the associated error. These 10000 random values per line are used to perform 10000 new calculations of physical conditions and ionic and total abundances. We started the propagation of errors with a lower number of Monte Carlo simulations and increased the number until the errors calculated for all quantities remained constant. The central 68 per cent of the distributions of values are used to provide the one-$\sigma$ uncertainties presented below.

\begin{table}
 \caption{Atomic data. } \label{table2}
 \centering
 \begin{tabular}{l l l}
  \hline
  Ion& Transition probabilities & Collision strengths\\
  \hline
  O$^{+}$ & \citeauthor*{FFT04}	& \citet{KSFK09} \\
  & \citeyearpar{FFT04} & \\
  O$^{++}$ & \citet{Storey00}$^\mathrm{a}$ & \citet{SSB14}\\
  & \citet{Wiese96} & \\
  N$^{+}$ & \citeauthor*{FFT04} & \citet{T11}\\
  & \citeyearpar{FFT04} & \\
  S$^{+}$ & \citet{FFTI06} & \citet{TZ10}\\
  S$^{++}$ & \citet{FFTI06} & \citet{GRHK14}\\
  Ne$^{++}$ &\citet{McL11} & \citet{McL11}\\
  Ne$^{3+}$ &\citet{GFF84} & \citet{G81}\\
  Ne$^{4+}$ &\citet{GMZ97} & \citet{DPNP13}\\
  & \citet{BD93}$^\mathrm{b}$ & \\
  Ar$^{++}$ & \citet{KS86} & \citet{GMZ95}\\
  &\citet{M83}$^\mathrm{c}$ & \\
  Ar$^{3+}$ & \citet{MZ82b} & \citet{RB97}\\
  Ar$^{4+}$ & \citet{LL93}$^\mathrm{d}$ & \citet{GMZ95}\\
  & \citet{MZ82b}$^\mathrm{e}$ & \\
  & \citet{KS86} & \\
  Cl$^{+}$ & \citet{MZ83} & \citet{T04}\\
  Cl$^{++}$ & \citet{Fal99}& \citet{BZ89}\\
  Cl$^{3+}$ & \citet{KS86}  & \citet{GMZ95}\\
  & \citet{EM84}$^\mathrm{f}$ & \\
  & \citet{MZ82a}$^\mathrm{g}$ & \\
  Fe$^{++}$ & \citet{Q96} & \citet{Z96}\\
  Fe$^{3+}$ &\citet{FFRR08} & \citet{ZP97}\\
  K$^{3+}$ & \citet{M83}$^\mathrm{h}$ & \citet{GMZ95}\\
  & \citet{KS86} & \\
  \hline
  \multicolumn{3}{l}{$^\mathrm{a}$Data for transitions 4--2 and 4--3.}\\
  \multicolumn{3}{l}{$^\mathrm{b}$Data for transitions 6--2 and 6--3.}\\
  \multicolumn{3}{l}{$^\mathrm{c}$Data for transitions 3--1, 4--3, and 5--1.}\\
  \multicolumn{3}{l}{$^\mathrm{d}$Data for transitions from level 6. }\\
  \multicolumn{3}{l}{$^\mathrm{e}$Data for transitions 3--1, 4--1 and 5--3. }\\
  \multicolumn{3}{l}{$^\mathrm{f}$Data for transitions 6--2 and 6--3.}\\
  \multicolumn{3}{l}{$^\mathrm{g}$Data for transitions 3--1, 4--1 and 5--3.}\\
  \multicolumn{3}{l}{$^\mathrm{h}$Data for transitions 3--1, 4--3 and 5--1.}\\
 \end{tabular}
\end{table}

\subsection{Physical conditions}\label{neTe}

We use an iterative procedure to derive the electron density, $n_{\text{e}}$, and electron temperature, $T_{\text{e}}$. We assume an initial $T_{\text{e}}$ of 10000~K and calculate $n_{\text{e}}$ from the available density diagnostics, [\ion{O}{ii}]~$\lambda3727/\lambda3729$, [\ion{S}{ii}]~$\lambda6716/\lambda6731$, [\ion{Cl}{iii}]~$\lambda5518/\lambda5538$, and [\ion{Ar}{iv}] $\lambda4711/\lambda4740$. Then, we calculate the mean of the individual logarithmic values of each diagnostic, excluding those values that differ by more than a factor of 2 from the median value. The mean value is used to compute $T_{\text{e}}$ for the different temperature diagnostics, [\ion{N}{ii}]~$(\lambda6548+\lambda6584)/\lambda5755$, [\ion{O}{ii}]~$(\lambda3727+\lambda3729)/(\lambda7319+\lambda7330)$, [\ion{O}{iii}]~$(\lambda4959+\lambda5007)/\lambda4363$, [\ion{S}{ii}]~$(\lambda6716+\lambda6731)/(\lambda4069+\lambda4076)$, [\ion{S}{iii}]~$\lambda9069/\lambda6312$, and [\ion{Ar}{iii}]~$\lambda7136/\lambda5192$. The values of $T_{\text{e}}$([\ion{N}{ii}]) are used to calculate new values of $n_{\text{e}}$([\ion{S}{ii}]), $n_{\text{e}}$([\ion{O}{ii}]), and $n_{\text{e}}$([\ion{Cl}{iii}]). The value of $T_{\text{e}}$([\ion{O}{iii}]) is used to derive a new value for $n_{\text{e}}$([\ion{Ar}{iv}]). The procedure is repeated until the values of $T_{\text{e}}$ and $n_{\text{e}}$ converge. 

We could not detect the [\ion{N}{ii}]~$\lambda5755$ line in the spectrum of N90. The value of $T_{\text{e}}$([\ion{N}{ii}]) is estimated for this object using the temperature relation provided by \citet{EC09}.

The [\ion{N}{ii}] and [\ion{O}{ii}] line intensities can be affected by recombination, which will lead to temperature values higher than the real ones. Following the procedure described by \citet{Rodriguez20}, we find that the corrections for $T_{\text{e}}$([\ion{O}{ii}]) are higher than 500 K for N44C and NGC~1714 and lower for the rest of the objects. The corrected and uncorrected values of  $T_{\text{e}}$([\ion{O}{ii}]) are both listed in Table \ref{table3}. The corrections for $T_{\text{e}}$([\ion{N}{ii}]) can be considered negligible, being lower than $\sim100$~K in all cases. 

We use $T_{\text{e}}$([\ion{N}{ii}]) and $T_{\text{e}}$([\ion{O}{iii}]) to characterize the gas of the nebula. We do not use the other available temperature diagnostics,  $T_{\text{e}}$([\ion{O}{ii}]), $T_{\text{e}}$([\ion{S}{iii}]), $T_{\text{e}}$([\ion{S}{ii}]), and $T_{\text{e}}$([\ion{Ar}{iii}]) because they are affected by different problems (recombination effects, telluric absorption, density variations) and/or have higher uncertainties than $T_{\text{e}}$([\ion{N}{ii}]) and $T_{\text{e}}$([\ion{O}{iii}]).

Table \ref{table3} presents the final values of electron density and temperature obtained from our spectra; the values derived with previous spectra from the literature are presented in Appendix~\ref{appendixB}.

\begin{table*}
 \centering
 \caption{Electron densities and temperatures of \ion{H}{ii} regions in the SMC and LMC.}
 \label{table3}
 \begin{tabular}{llcccc} 
  \hline
 Diagnostic & Intensity line ratio & IC~2111 & N11B &  N44C & NGC~1714\\
 & & (LMC) & (LMC) &  (LMC) & (LMC)\\
  \hline
  $n_\text{e}$ (cm$^{-3}$)& [\ion{O}{ii}] $\lambda3727/\lambda3729$ & $300^{+140}_{-110}$ & $270\pm100$ & $190\pm90$ & $460^{+170}_{-140}$ \\
  & [\ion{S}{ii}] $\lambda6716/\lambda6731$ & $290^{+140}_{-120}$ & $310\pm80$  & $120\pm60$ & $440^{+170}_{-140}$ \\
  & [\ion{Cl}{iii}] $\lambda5518/\lambda5538$ & $700^{+630}_{-390}$  & $340\pm150$ & $520^{+300}_{-260}$ & $400^{+680}_{-220}$ \\
  & [\ion{Ar}{iv}] $\lambda4711/\lambda4740$ & $1150^{+4150}_{-460}$  & $610^{+680}_{-340}$ & $540^{+460}_{-300}$ & $1180^{+1880}_{-670}$ \\
  & Adopted value & $390\pm130$  & $360\pm90$ & $360\pm130$ & $460\pm160$\\
  \multicolumn{6}{c}{}\\
  $T_\text{e}$ (K)& [\ion{O}{ii}] $(\lambda3727+\lambda3729)/(\lambda7319+\lambda7330)$ & $9380^{+590}_{-480}$ & $9970^{+480}_{-400}$ & $11870^{+910}_{-760}$ & $10280^{+790}_{-650}$ \\
  & [\ion{O}{ii}] $(\lambda3727+\lambda3729)/(\lambda7319+\lambda7330)^\mathrm{a}$ & $9140^{+590}_{-480}$ & $9750^{+480}_{-400}$ & $11200^{+910}_{-760}$ & $9720^{+790}_{-650}$ \\
  & [\ion{O}{iii}] $(\lambda4959+\lambda5007)/\lambda4363$ & $9130\pm150$ & $9160\pm100$ & $11310\pm150$ & $9530\pm150$ \\
  & [\ion{N}{ii}] $(\lambda6548+\lambda6584)/\lambda5755$ & $9780\pm340$ & $9800\pm120$ & $10510\pm330$ & $10240^{+540}_{-560}$\\
  & [\ion{S}{ii}] $(\lambda6716+\lambda6731)/(\lambda4069+\lambda4076)$ & $9840^{+840}_{-720}$  & $10790^{+730}_{-640}$ & $10630^{+960}_{-820}$ & $11500^{+1300}_{-1100}$\\
  & [\ion{S}{iii}] $\lambda9069/\lambda6312$ & $8040^{+270}_{-240}$ & $9970^{+260}_{-230}$ & $11140^{+330}_{-300}$ & $8140\pm270$ \\
  & [\ion{Ar}{iii}] $\lambda7136/\lambda5192$ & $9010^{+710}_{-770}$ & $9730\pm250$ & $10650\pm520$ & $9390^{+730}_{-750}$ \\
  \hline
  Diagnostic & Intensity line ratio & N66A & N81 & N88A & N90\\
  & & (SMC) & (SMC) &  (SMC) & (SMC)\\
  \hline
  $n_\text{e}$ (cm$^{-3}$)& [\ion{O}{ii}] $\lambda3727/\lambda3729$ & $190^{+110}_{-90}$ & $440^{+150}_{-130}$ & $2550^{+570}_{-440}$ & $90^{+510}_{-20}$ \\
  & [\ion{S}{ii}] $\lambda6716/\lambda6731$ & $210\pm80$ & $380\pm110$ & $2110^{+450}_{-370}$ & $130^{+110}_{-70}$ \\
  & [\ion{Cl}{iii}] $\lambda5518/\lambda5538$ & $390^{+250}_{-210}$ & $470\pm180$ & $3770\pm300$ & -- \\
  & [\ion{Ar}{iv}] $\lambda4711/\lambda4740$ & -- & $970^{+490}_{-430}$ & $5720^{+830}_{-760}$ & -- \\
  & Adopted value & $230\pm80$ & $420\pm90$ & $3280^{+280}_{-250}$ & $170^{+140}_{-90}$ \\
  \multicolumn{6}{c}{}\\
$T_\text{e}$ (K)  & [\ion{O}{ii}] $(\lambda3727+\lambda3729)/(\lambda7319+\lambda7330)$ & $12900^{+780}_{-650}$ & $12700^{+740}_{-630}$ & $13990^{+920}_{-820}$ & $11900^{+1300}_{-1000}$ \\
  & [\ion{O}{ii}] ($\lambda3727+\lambda3729)/(\lambda7319+\lambda7330)^\mathrm{a}$ & $12750^{+780}_{-650}$ & $12480^{+740}_{-630}$ & $13630^{+920}_{-820}$ & $11800^{+1300}_{-1000}$ \\
  & [\ion{O}{iii}] $(\lambda4959+\lambda5007)/\lambda4363$ & $12540\pm190$ & $12830\pm200$ & $13920\pm230$ & $12100^{+180}_{-200}$ \\
  & [\ion{N}{ii}] $(\lambda6548+\lambda6584)/\lambda5755$ & $12080\pm320$ & $11900\pm270$ & $13120\pm260$ & $11640\pm140^\mathrm{b}$ \\
  & [\ion{S}{ii}] $(\lambda6716+\lambda6731)/(\lambda4069+\lambda4076)$ & $19400^{+2350}_{-1960}$ & $13440^{+1060}_{-930}$ & $12650^{+1060}_{-930}$ & $12900\pm1300$ \\
  & [\ion{S}{iii}] $\lambda9069/\lambda6312$ & $14030^{+600}_{-540}$ & $13740^{+630}_{-550}$ & $16430^{+930}_{-810}$ & $13240^{+1100}_{-950}$ \\
  & [\ion{Ar}{iii}] $\lambda7136/\lambda5192$ & $12200\pm620$ & $11880\pm280$ & $12930^{+300}_{-270}$ & -- \\
  \hline
  \multicolumn{6}{l}{$^\mathrm{a}$Values of $T_\text{e}$([\ion{O}{ii}]) corrected for recombination effects.}\\
  \multicolumn{6}{l}{$^\mathrm{b}$$T_\text{e}$([\ion{N}{ii}]) obtained with the empirical temperature relation given by \citet{EC09}.}\\
 \end{tabular}
\end{table*}

\subsection{Ionic and total abundances}

We compute the ionic abundances using collisionally excited lines for all the available ions where $T_\text{e}$([\ion{N}{ii}]) is used for O$^+$, N$^+$, S$^+$, and Fe$^{++}$, $T_\text{e}$([\ion{O}{iii}]) is used for O$^{++}$, Ne$^{++}$, He$^+$, and He$^{++}$, and the mean of $T_{\text{e}}$([\ion{N}{ii}]) and $T_{\text{e}}$([\ion{O}{iii}]) is used for S$^{++}$, Ar$^{++}$, and Cl$^{++}$ \citep[][Dom\'inguez-Guzm\'an et al. 2022, in prep.]{DG19}.

The emission lines that are used to calculate the ionic abundances are:  \ion{He}{I} $\lambda4471$, $\lambda5876$, $\lambda6678$, \ion{He}{ii} $\lambda4686$, [\ion{O}{ii}] $\lambda\lambda3726$, 3729, [\ion{O}{iii}] $\lambda\lambda4959$, 5007, [\ion{N}{ii}] $\lambda\lambda6548$, 6584, [\ion{S}{ii}] $\lambda\lambda6716$, 6731, [\ion{S}{iii}] $\lambda6312$, [\ion{Ne}{iii}] $\lambda\lambda3869$, 3968, [\ion{Ne}{iv}] $\lambda\lambda4724$, 4726, [\ion{Ne}{v}] $\lambda3426$, [\ion{Ar}{iii}] $\lambda7136$, [\ion{Ar}{iv}] $\lambda\lambda4711$, 4740, [\ion{Ar}{v}] $\lambda6435$, [\ion{Cl}{ii}] $\lambda9124$, [\ion{Cl}{iii}] $\lambda\lambda5518$, 5538, [\ion{Cl}{iv}] $\lambda\lambda7531$, 8046, [\ion{Fe}{iv}] $\lambda6740$, [\ion{K}{iv}] $\lambda\lambda6102$, 6796.

In the case of Fe$^{++}$, we use the following emission lines: [\ion{Fe}{iii}] $\lambda4009$, $\lambda4659$, $\lambda4701$, $\lambda4734$, $\lambda4755$, $\lambda4770$, $\lambda4778$, $\lambda4881$, $\lambda4986$, $\lambda4987$, $\lambda5270$, and $\lambda5412$. The [\ion{Fe}{iii}] $\lambda4607$ line is not used because it is blended with a \ion{N}{ii} line. On the other hand, [\ion{Fe}{iii}] $\lambda4667$ might be blended with an unidentified feature since it leads to Fe$^{++}$ abundances that are 0.2--1.3~dex higher than those implied by other [\ion{Fe}{iii}] lines. Besides, [\ion{Fe}{iii}] $\lambda\lambda3240$, 3323 are measured in N88A but not used because the flux calibration is very uncertain below 3300 \AA, where the first transition lies, and because the Einstein coefficient is not available for the second transition.

When several emission lines of the same ion are available, a sum of the line intensities is taken, excluding those lines that are blended with others or affected by sky features. The values of the ionic abundances derived with our spectra are presented in Table \ref{table4}; the ionic abundances obtained with the previously available spectra are presented in Appendix~\ref{appendixB}.

\begin{table*}
 \centering
 \caption{Ionic abundances in units of 12+log(X$^{+i}$/H$^+$) of \ion{H}{ii} regions in the LMC and SMC.}
 \label{table4}
 \begin{tabular}{l c c c c c c c c}
  \hline
  & IC~2111 & N11B & N44C & NGC~1714 & N66A & N81 & N88A & N90  \\
  Ion & (LMC) & (LMC) & (LMC) & (LMC) & (SMC) & (SMC) & (SMC) & (SMC)  \\
  \hline
  O$^{+}$ & $8.07\pm0.07$ &  $7.99\pm0.03$ & $7.33\pm0.06$ & $7.66^{+0.11}_{-0.09}$ & $7.49\pm0.04$ & $7.30\pm0.04$ & $6.88\pm0.04$ & $7.70\pm0.06$ \\
  O$^{++}$ & $8.18\pm0.04$ & $8.17\pm0.02$ & $8.22\pm0.02$ & $8.27\pm0.04$ & $7.84\pm0.02$ & $7.90\pm0.02$ & $7.98\pm0.02$ & $7.79\pm0.02$ \\
  N$^{+}$ & $6.66\pm0.05$ & $6.62\pm0.02$ & $6.04\pm0.04$ & $6.25\pm0.07$ & $5.95\pm0.03$ & $5.70\pm0.03$ & $5.41\pm0.03$ & $6.13\pm0.03$ \\
  S$^{+}$ & $5.84\pm0.04$ & $5.81\pm0.02$ & $5.56\pm0.04$ & $5.47\pm0.06$ & $5.46\pm0.03$ & $5.17\pm0.03$ & $4.98\pm0.03$ & $5.64\pm0.02$ \\
  S$^{++}$ & $6.70\pm0.05$ & $6.65\pm0.02$ & $6.34\pm0.03$ & $6.60\pm0.07$ & $6.27\pm0.03$ & $6.26\pm0.03$ & $6.14\pm0.02$ & $6.28\pm0.05$ \\
  Ne$^{++}$ & $7.54\pm0.04$ & $7.57\pm0.02$ & $7.68\pm0.03$ & $7.70\pm0.04$ & $7.24\pm0.03$ & $7.34\pm0.03$ & $7.39\pm0.03$ & $7.16\pm0.03$ \\
  Ne$^{3+}$ & -- & -- & $7.40\pm0.06$ & -- & -- & -- & -- & -- \\
  Ne$^{4+}$ & -- & -- & $5.32\pm0.05$ & -- & -- & -- & -- & -- \\
  Ar$^{++}$ & $6.01\pm0.03$ & $5.98\pm0.02$ & $5.80\pm0.02$ & $5.99\pm0.04$ & $5.66\pm0.02$ & $5.71\pm0.02$ & $5.63\pm0.02$ & $5.68\pm0.03$ \\
  Ar$^{3+}$ & $4.17\pm0.08$ & $4.28\pm0.02$ & $5.74\pm0.02$ & $4.54\pm0.05$ & $4.32\pm0.02$ & $4.43\pm0.02$ & $4.95\pm0.02$ & -- \\
  Ar$^{4+}$ & -- & -- & $4.65\pm0.05$ & -- & -- & -- & -- & -- \\
  Cl$^{+}$ & $3.66^{+0.14}_{-0.13}$ & $3.72\pm0.02$ & $3.36\pm0.04$ & -- & -- & -- & -- & -- \\
  Cl$^{++}$ & $4.74\pm0.04$ & $4.72\pm0.01$ & $4.45\pm0.02$ & $4.68\pm0.05$ & $4.31\pm0.02$ & $4.35\pm0.02$ & $4.25\pm0.02$ & -- \\
  Cl$^{3+}$ & -- & -- & $4.45\pm0.02$ & $3.44\pm0.09$ & $3.28\pm0.03$ & $3.33\pm0.02$ & $3.79\pm0.02$ & -- \\
  Fe$^{++}$ & $5.22\pm0.06$ & $5.09\pm0.02$ & $4.43^{+0.06}_{-0.04}$ & $5.06\pm0.08$ & $4.82\pm0.04$ & $4.87\pm0.03$ & $4.98\pm0.02$ & $<4.38$ \\
  Fe$^{3+}$ & -- & -- & -- & -- & -- & -- & $5.46\pm0.04$ & -- \\
  K$^{3+}$ & -- & -- & $3.96\pm0.04$ & -- & -- & -- & -- & -- \\
  He$^{+}$ & $10.92\pm0.01$ & $10.91\pm0.01$ & $10.85\pm0.01$ & $10.92\pm0.01$ & $10.89\pm0.01$ & $10.89\pm0.01$ & $10.88\pm0.01$ & $10.92\pm0.02$ \\
  He$^{++}$ & -- & -- & $10.10\pm0.02$ & -- & -- & -- & -- & -- \\
  \hline
 \end{tabular}
\end{table*}

The corrections for the effects of recombination in [\ion{O}{ii}] and [\ion{N}{ii}] emission lines, which have been estimated as described above in Section~\ref{neTe}, can be considered negligible. The N$^+$ abundances and the O$^+$ abundances derived with the blue [\ion{O}{ii}] lines change by less than 0.01~dex. The corrections in the O$^+$ abundances derived with the red [\ion{O}{ii}] lines go up to $-0.04$~dex in N44C, but the results implied by the blue [\ion{O}{ii}] lines are the ones used to calculate all the final abundances.

The total oxygen abundance is calculated by adding O$^+$/H$^+$ and O$^{++}$/H$^+$, except for N44C, where \ion{He}{ii} emission is observed and we use the ICF given by \citet{Izo99}, which is based on the relative abundances determined for He$^+$ and He$^{++}$.

The total abundances of N, S, Ne, and Ar are calculated using the ionization correction factors (ICFs) given by \citet{ADIS21}, and the errors they provide are quadratically added to the uncertainties obtained from the Monte Carlo simulations. These ICFs correct for the presence of the unobserved ionization stages of these elements using fits to the results obtained from grids of photoionization models.

The total abundance of chlorine is obtained by adding the Cl$^{++}$ and Cl$^{3+}$ ionic abundances or using the empirical ICF of Dom\'inguez-Guzm\'an et al. (2022, in prep.)\footnote{For $x=\log(\mbox{O}^{++}/\mbox{O}^+)$, if $x\le0.2$, Cl/O\ $=(\mbox{Cl}^{++}/\mbox{H}^+)/(\mbox{O}/\mbox{H})$; if $0.2>x>1.1$, Cl/O\ $=(\mbox{Cl}^{++}/\mbox{O}^+)10^{-0.2419-0.7178x}$.} when the Cl$^{3+}$ abundance is not available. The differences between the chlorine abundances obtained in this way and those implied by the ICF of \citet{ADIS21} are lower than 0.1~dex.

The ICFs used for Fe are those prescribed by \citet{RR05} because they give us extreme values of the total iron abundance that can be used to constrain the true values of the Fe abundances in the gas. All these ICFs are based on the ratio of the O$^+$ and O$^{++}$ ionic abundances.

The total abundances for all the elements are presented in Table~\ref{table5} for the objects in our main sample; the results for the extended sample are listed in Appendix~\ref{appendixB}.

 \begin{table*}
 \centering
 \caption{Total abundances in units of 12+log(X/H).}
 \label{table5}
 \begin{tabular}{l c c c c c c c c}
  \hline
  & IC~2111 & N11B & N44C & NGC~1714 & N66A & N81 & N88A & N90  \\
   & (LMC) & (LMC) & (LMC) & (LMC) & (SMC) & (SMC) & (SMC) & (SMC)  \\
  \hline
  O & $8.43\pm0.04$ &  $8.39\pm0.02$ & $8.34\pm0.02$ & $8.37\pm0.04$ & $8.00\pm0.02$ & $8.00\pm0.02$ & $8.02\pm0.02$ & $8.05\pm0.03$ \\
  N & $7.09^{+0.16}_{-0.08}$ & $7.09^{+0.17}_{-0.07}$ & $7.15^{+0.19}_{-0.10}$ & $7.03^{+0.20}_{-0.08}$ & $6.53^{+0.19}_{-0.07}$ & $6.48^{+0.20}_{-0.08}$ & $6.64^{+0.18}_{-0.11}$ & $6.55^{+0.16}_{-0.08}$ \\
  S & $6.76\pm0.04$ & $6.72\pm0.02$ & $6.58^{+0.07}_{-0.05}$ & $6.67\pm0.06$ & $6.35\pm0.03$ & $6.34\pm0.03$ & $6.33^{+0.10}_{-0.07}$ & $6.38\pm0.04$ \\
  Ne & $7.86^{+0.16}_{-0.20}$ & $7.85^{+0.15}_{-0.18}$ & $7.83\pm0.05$ & $7.82^{+0.10}_{-0.08}$ & $7.45\pm0.14$ & $7.47\pm0.09$ & $7.44\pm0.04$ & $7.50^{+0.16}_{-0.20}$ \\
  Ar & $6.05\pm0.05$ & $6.02\pm0.04$ & $5.95^{+0.07}_{-0.05}$ & $6.03\pm0.05$ & $5.69\pm0.03$ & $5.75\pm0.04$ & $5.75^{+0.11}_{-0.07}$ & $5.72\pm0.04$ \\
  Cl & $4.78\pm0.04$ & $4.75\pm0.01$ & $4.58\pm0.02$ & $4.70\pm0.04$ & $4.33\pm0.02$ & $4.37\pm0.02$ & $4.36\pm0.02$ & -- \\
  Cl$^\mathrm{a}$ & -- & -- & $4.75\pm0.02$ & $4.70\pm0.04$ & $4.35\pm0.02$ & $4.39\pm0.02$ & $4.38\pm0.01$ & -- \\
  Fe$^\mathrm{b}$ & $5.56\pm0.06$ & $5.43\pm0.02$ & $4.97\pm0.04$ & $5.45\pm0.06$ & $5.17\pm0.03$ & $5.26\pm0.02$ & $5.52\pm0.02$ & $<4.72$ \\
  Fe$^\mathrm{c}$ & $5.53\pm0.03$ & $5.43\pm0.02$ & $5.32\pm0.04$ & $5.67\pm0.04$ & $5.26\pm0.02$ & $5.48\pm0.02$ & $5.99\pm0.03$ & $<4.68$ \\
  Fe$^\mathrm{d}$ & -- & -- & -- & -- & -- & -- & $5.59\pm0.03$& -- \\
  He$^\mathrm{e}$ & $10.92\pm0.01$ & $10.91\pm0.01$ & $10.92\pm0.01$ & $10.92\pm0.01$ & $10.89\pm0.01$ & $10.89\pm0.01$ & $10.88\pm0.01$ & $10.92\pm0.02$ \\
  \hline
  \multicolumn{9}{l}{$^\mathrm{a}$Total chlorine abundance obtained adding the ionic abundances of Cl$^{++}$ and Cl$^{3+}$.}\\
  \multicolumn{9}{l}{$^\mathrm{b}$Derived using equation (3) of \citet{RR05}.}\\
  \multicolumn{9}{l}{$^\mathrm{c}$Derived using equation (2) of \citet{RR05}.}\\
  \multicolumn{9}{l}{$^\mathrm{d}$Total iron abundance obtained adding the ionic abundances of Fe$^{++}$ and Fe$^{3+}$.}\\
  \multicolumn{9}{l}{$^\mathrm{e}$Lower limit to the total helium abundance if He$^0$ has a significant concentration.}\\
 \end{tabular}
\end{table*}

\section{Results and discussion}\label{sec6}

Table~\ref{table6} shows the proto-solar abundances of \citet{Lodd19} along with the weighted means and standard deviations, $\sigma$, of different abundance ratios obtained either with the UVES/VLT data or with previous spectra. Note that we are providing the standard deviations but not the standard errors (the errors of the mean) because we want to quantify the size of the variations. We do not include N44C in these calculations for the reasons stated above. The results are discussed in the following subsections.
 
 \begin{table*}
 \centering
 \caption{Weighted means and standard deviations of the X/H and X/O abundance ratios, in units of $12+\log(\mbox{X/H})$ and $\log(\mbox{X/O})$, respectively, obtained with the UVES/VLT spectra and with other spectra. In the LMC we include 30~Dor but not N44C (see text).}
 \label{table6}
 \begin{tabular}{l c c c c c c}
  \hline
  & Proto-solar & Galaxy & \multicolumn{2}{c}{Previous spectra} & \multicolumn{2}{c}{UVES/VLT spectra} \\
  & abundances$^\mathrm{a}$& &  Mean  & $\sigma$  &  Mean  & $\sigma$  \\
  \hline
   He/H & $10.994\pm0.02$ & LMC & $10.93$ & $0.04$ & $10.93$ & $0.01$ \\ 
  & & SMC & $10.91$ & $0.04$  & $10.89$ & $0.02$ \\ 
  O/H & $8.82\pm0.07$ & LMC & $8.36$ & $0.07$  & $8.37$ & $0.03$ \\ 
  & & SMC & $8.03$ & $0.06$ & $8.01$ & $0.02$ \\ 
  N/H & $7.94\pm0.12$ & LMC & $7.08$ & $0.18$ & $7.09$ & $0.05$ \\ 
  & & SMC & $6.72$ & $0.22$ & $6.55$ & $0.07$ \\ 
  S/H & $7.24\pm0.03$ & LMC & $6.67$ & $0.13$  & $6.72$ & $0.04$ \\ 
  & & SMC & $6.33$ & $0.08$ & $6.35$ & $0.02$ \\ 
  Cl/H & $5.32\pm0.06$ & LMC & $4.87$ & $0.21$ & $4.73$ & $0.04$ \\ 
  & & SMC & $4.57$ & $0.21$ & $4.36$ & $0.02$ \\ 
  Ar/H & $6.59\pm0.10$ & LMC & $6.07$ & $0.07$  & $6.04$ & $0.02$ \\ 
   & & SMC & $5.70$ & $0.04$  & $5.72$ & $0.03$ \\ 
  Ne/H & $8.24\pm0.10$ & LMC & $7.79$ & $0.06$ & $7.82$ & $0.03$ \\ 
   & & SMC & $7.49$ & $0.09$  & $7.44$ & $0.03$ \\ 
  Fe/H  & $7.54\pm0.02$ & LMC & $5.67^\mathrm{b}/5.82^\mathrm{c}$ & $0.12^\mathrm{b}/0.22^\mathrm{c}$ & $5.52^\mathrm{b}/5.75^\mathrm{c}$ & $0.11^\mathrm{b}/0.21^\mathrm{c}$ \\ 
   & & SMC & $5.51^\mathrm{b}/5.74^\mathrm{c}$ & $0.37^\mathrm{b}/0.49^\mathrm{c}$  & $5.40^\mathrm{b}/5.54^\mathrm{c}$ & $0.18^\mathrm{b}/0.38^\mathrm{c}$ \\ 
  N/O & $-0.88\pm0.14$ & LMC & $-1.29$ & $0.20$ & $-1.30$ & $0.07$ \\ 
  & & SMC & $-1.32$ & $0.24$ & $-1.47$ & $0.06$ \\ 
  S/O & $-1.58\pm0.08$ & MCs & $-1.70$ & $0.08$  & $-1.66$ & $0.02$ \\ 
  Cl/O & $-3.50\pm0.09$ & MCs & $-3.46$ & $0.16$  & $-3.65$ & $0.02$ \\ 
  Ar/O & $-2.23\pm0.12$ & MCs & $-2.32$ & $0.08$  & $-2.32$ & $0.05$ \\ 
  Ne/O & $-0.58\pm0.12$ & MCs & $-0.52$ & $0.07$ & $-0.57$ & $0.02$ \\ 
  Fe/O & $-1.28\pm0.07$ & MCs & $-2.60^\mathrm{b}/-2.50^\mathrm{c}$ & $0.29^\mathrm{b}/0.46^\mathrm{c}$ & $-2.73^\mathrm{b}/-2.57^\mathrm{c}$ & $0.16^\mathrm{b}/0.32^\mathrm{c}$ \\ 
  \hline
  \multicolumn{7}{l}{$^\mathrm{a}$\citet{Lodd19}.}\\
  \multicolumn{7}{l}{$^\mathrm{b}$Derived using equation (3) of \citet{RR05}.}\\
  \multicolumn{7}{l}{$^\mathrm{c}$Derived using equation (2) of \citet{RR05}.}\\
 \end{tabular}
\end{table*}

\subsection{Comparison with previous spectra}\label{subsec1}

The results presented in Table~\ref{table6} show that most of the mean values of the abundance ratios do not change much when the UVES/VLT spectra are used instead of the lower-quality spectra. The exceptions are the Cl/H and Cl/O abundance ratios, where the differences in the mean values go up to 0.22 dex, probably because of the faintness of the [\ion{Cl}{iii}] lines. On the other hand, the standard deviations are clearly much larger when calculated with the previous spectra, showing that the deep UVES/VLT spectra provide much better estimates of the chemical abundances and their variations. This comparison illustrates that the deep UVES/VLT spectra provide estimates of the chemical abundances of the ISM and their spatial variations in the MCs with unprecedented quality and precision. Moreover, the remarkably small standard deviations of the total abundances determined in the LMC and SMC further support the absence of significant chemical inhomogeneities in the ISM inside each one of the MCs.

\subsection{The He/H abundance ratio}\label{subsec2}

The mean values of the helium abundances in the MCs implied by the UVES/VLT spectra are $12+\log(\mbox{He}/\mbox{H})= 10.89$ (SMC) and $12+\log(\mbox{He}/\mbox{H})= 10.93$ (LMC). The He/H abundance is lower in the SMC than in the LMC by 0.04~dex,  and lower in the LMC than in the protosun by 0.06~dex. These differences are compatible with the expected increase of the helium abundance with metallicity \citep[see, e.g.,][]{Izotov07,Peimbert07,Vital18}. However, there are variations in the values of He/H from H~{\sc ii} region to H~{\sc ii} region inside each MC, but we cannot ascertain if these variations are due to failures in our assumption that $\mbox{He}/\mbox{H}\simeq\mbox{He}^+/\mbox{H}^+ + \mbox{He}^{++}/\mbox{H}^+$.

 \subsection{The O/H abundance ratio}\label{subsec3}
 
\begin{figure}
	\center
	\includegraphics[width=0.42\textwidth, trim=10 7 5 5, clip=yes]{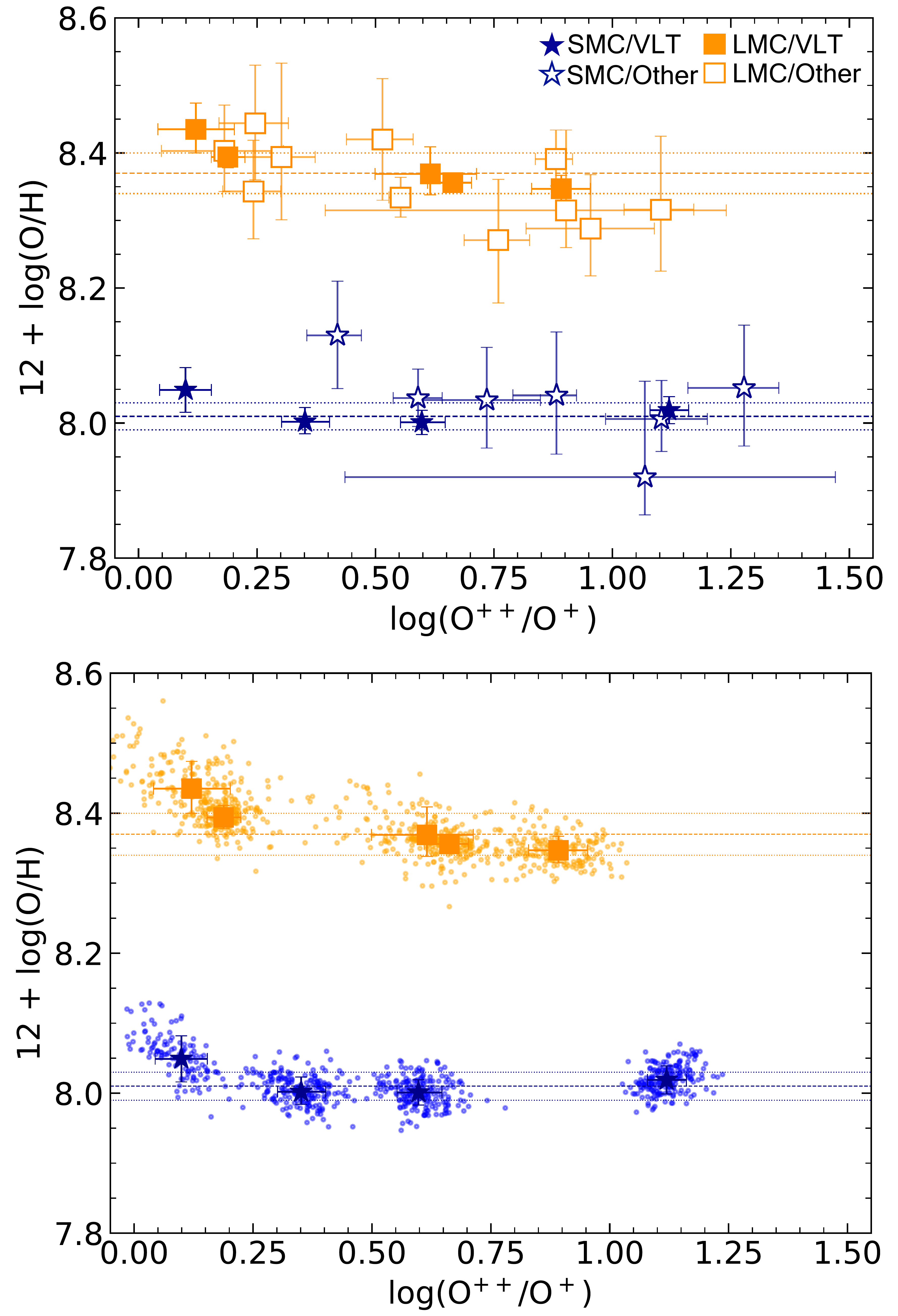}
	\caption{Oxygen abundances of MC \ion{H}{ii} regions as a function of the degree of ionization. The squares represent regions in the LMC, and the stars regions in the SMC. In the upper panel, filled symbols show our results for the UVES/VLT observations and empty symbols the results we obtain from other spectra collected from the literature. The long dashed lines show the weighted mean oxygen abundances in the LMC and the SMC; the small dashed lines show the standard deviations. The lower panel shows with small dots the values of O/H and O$^{++}$/O$^{+}$ implied by 200 runs of the Monte Carlo propagation of errors for each UVES/VLT spectra.}
	\label{fig7}
\end{figure}

The upper panel of Fig.~\ref{fig7} shows the oxygen abundance of the sample objects as a function of the degree of ionization. The filled symbols are the values for the UVES/VLT spectra (our sample plus 30~Dor) and the empty ones are for the previous spectra. The squares are H~{\sc ii} regions from the LMC and the stars are H~{\sc ii} regions from the SMC. The long and small dashed lines show the weighted means and standard deviations of the oxygen abundances in the SMC and the LMC for the values calculated with the UVES/VLT spectra: $12+\log(\mbox{O}/\mbox{H})= 8.01$ (SMC) and $12+\log(\mbox{O}/\mbox{H})= 8.37$ (LMC). The O/H abundance is higher in the LMC than in the SMC by 0.36~dex for the results calculated with the UVES/VLT spectra and 0.33~dex for the previous spectra. The standard deviations obtained with the UVES/VLT spectra are equal to 0.02--0.03~dex, whereas the previous spectra imply standard deviations of 0.06--0.07~dex in both galaxies.

The oxygen abundances derived for the MCs are lower than the proto-solar abundance of \citet{Lodd19}, $12+\log(\mbox{O/H})=8.82\pm0.07$, by factors of $\sim3$ (LMC) and $\sim6.5$ (SMC). The nebular abundances should be corrected upwards by including the amount of oxygen deposited in dust grains, but the correction is probably smaller than 0.1~dex \citep{Peimbert10}. If we take the Orion nebula as representative of the interstellar abundance in the solar neighbourhood, its oxygen abundance, $12+\log(\mbox{O/H})=8.51\pm0.03$ \citep{Arellano20}, is higher than the MCs abundances by factors of 1.4 and 3. 

It can be seen in Fig.~\ref{fig7} that the oxygen abundances show slight trends with the degree of ionization. However, since both O/H and O$^{++}$/O$^{+}$ depend on the O$^{+}$ and O$^{++}$ abundances, errors in these ionic abundances can introduce spurious correlations between the two variables. In order to illustrate this, we show with small dots in the lower panel of Fig.~\ref{fig7} the results obtained with the UVES/VLT spectra for 200 of the 10000 results calculated for each object with the Monte Carlo runs used in the propagation of errors. In those objects with lower degree of ionization, these small dots define trends of O/H decreasing with O$^{++}$/O$^{+}$ which can explain most of the correlations. The trends are mostly due to changes in $T_{\text{e}}$([\ion{N}{ii}]) that affect the values of O$^{+}$/H$^{+}$. In fact, a similar and much more extreme trend of O/H with O$^{++}$/(O$^{+}+$~O$^{++}$) was found in the LMC by \citet{PEFW78}, who attributed it to problems with their assumed temperature structure.

The two objects with the highest values of O/H, N90 in the SMC and IC~2111 in the LMC, would require their values of $T_{\text{e}}$([\ion{N}{ii}]) to increase by $\sim400$ and 1000~K, respectively, in order to achieve the mean value O/H derived in each MC. In the case of N90, the value of $T_{\text{e}}$([\ion{N}{ii}]) was estimated from the value of $T_{\text{e}}$([\ion{O}{iii}]) using the temperature relation determined by \citet{EC09} and thus the uncertainty in $T_{\text{e}}$([\ion{N}{ii}]) is likely to be much higher than the value given in Table~\ref{table3}. In the case of IC~2111, the required change in $T_{\text{e}}$([\ion{N}{ii}]) is larger, and only three per cent of the values obtained with the Monte Carlo simulations for this object have O/H equal or lower than the mean value in the LMC. This difference might be real or there could be another unidentified source of uncertainty. On the other hand, the region with the lowest value of O/H in the LMC, N44C, is the only object that requires the use of an ICF for oxygen and this implies an additional source of uncertainty. Taking all these considerations into account, we conclude that the uncertainties in temperature and the ICF in N44C are responsible for part of the dispersion in the results. Hence, the intrinsic standard deviations of O/H in each MC should be lower than the values estimated from our results, 0.02--0.03~dex.

\subsection{Comparison with the abundances of young stars}\label{subsec3b}

Our chemical abundances can also be compared with those obtained for young OB-type stars. The abundances of oxygen and nitrogen have been calculated for many OB-type stars in the Magellanic Clouds \citep[and references therein]{Rolle02,Hunter07,Dufton20,Bouret21}, but the results cover wide ranges in the abundances of these elements. In the case of oxygen, the ranges are $12+\log(\mbox{O/H})=$\ 7.4--8.3 (SMC) and 8.1--8.5 (LMC), and our estimates fall within these ranges, with $12+\log(\mbox{O/H})=$\ 8.01 (SMC) and 8.37 (LMC). For nitrogen, the ranges of stellar abundances are $12+\log(\mbox{N/H})=$\ 6.3--8.4 (SMC) and 6.9--8.2 (LMC). The lower values of N/H for the SMC are upper limits, and our estimates of $12+\log(\mbox{N/H})=$\ 6.55 (SMC) and 7.14 (LMC) fall within these ranges of stellar abundances. 

The large dispersions in the abundances of oxygen and nitrogen in young massive stars of the MCs are usually attributed to the mixing of nuclear processed material, which is expected to lead to an enrichment in nitrogen and a deficit of oxygen \citep[see, e.g.,][]{Maeder14}. On the other hand, \citet{Przy08} obtained homogeneous chemical abundances in young stars of the solar neighbourhood by restricting their sample to unevolved early B-type stars with low rotational velocities. However, if we take the stars from \citet{Hunter07} for both MCs that have similar characteristics to those of \citet{Przy08} in terms of their spectral types and rotational velocities, the abundances of O and N still cover large ranges.  A possible explanation is that the stellar spectra in the MCs have lower quality than those used by \citet{Przy08} for stars in the solar neighbourhood.

 \subsection{The S/O, Ne/O, Ar/O, and Cl/O abundance ratios}\label{subsec4}
 
\begin{figure*}
	\center
	\includegraphics[width=0.8\textwidth, trim=10 7 5 5, clip=yes]{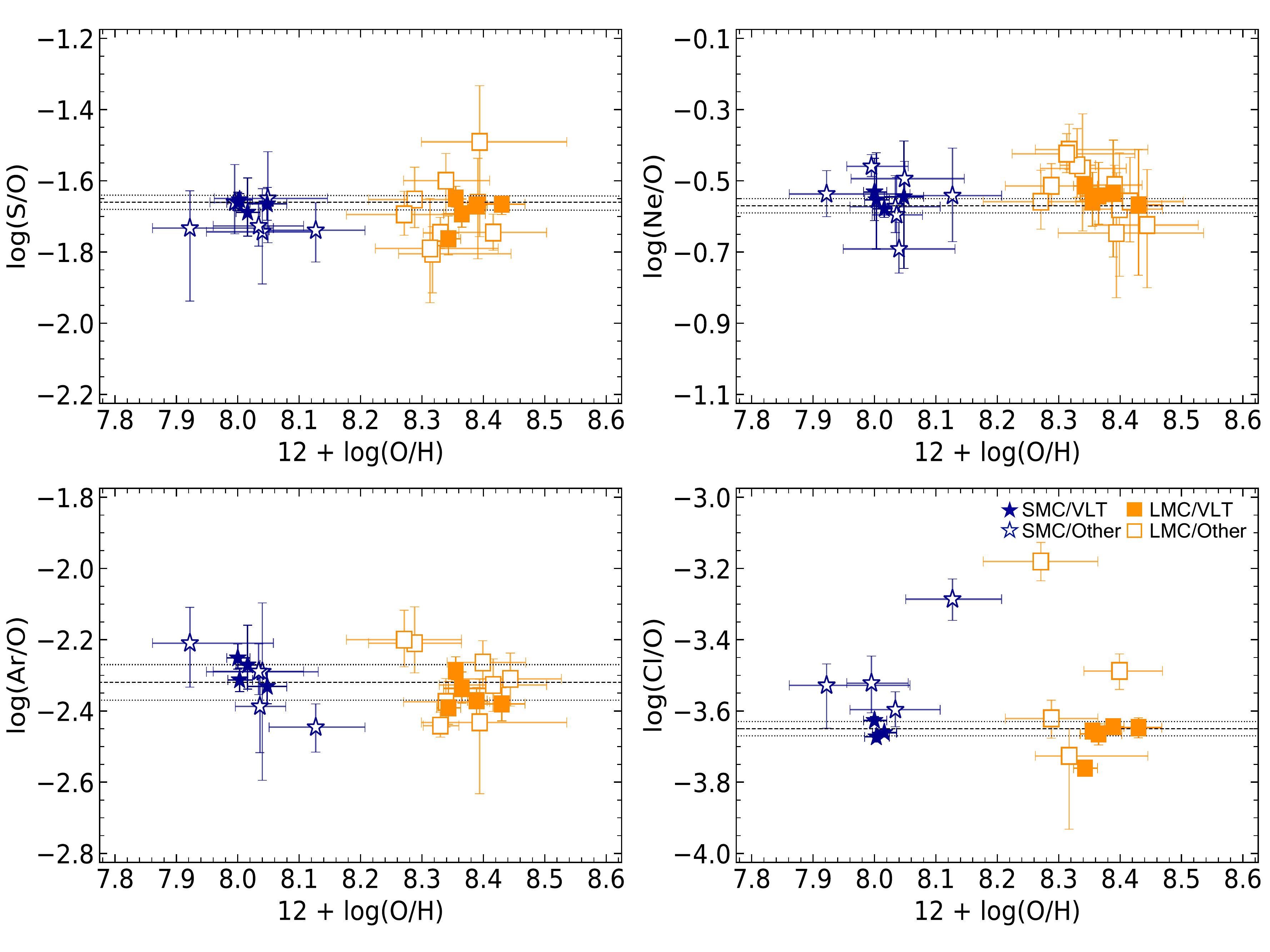}
	\caption{S/O, Ne/O, Ar/O, and Cl/O abundance ratios as a function of O/H. Squares and stars are \ion{H}{ii} regions in the LMC and the SMC, respectively. The filled symbols represent the results based on UVES/VLT observations; the empty ones are based on spectra gathered from the literature. The long and small dashed lines are the weighted means and standard deviations for the UVES/VLT spectra (excluding N44C). The graphs have the same vertical range in dex.}
	\label{fig8}
\end{figure*}

The values of the S/O, Ne/O, Ar/O, and Cl/O abundance ratios are shown as a function of the oxygen abundance in Fig.~\ref{fig8}. As  in the previous figure, we use squares and stars to show the results obtained for the LMC and SMC, respectively. Filled symbols represent the results derived with the UVES/VLT spectra; open symbols the results implied by other spectra from the literature. Long and small dashed lines show the values of the mean and standard deviation of each abundance ratio calculated with the UVES/VLT data. In order to better compare the variations between the different abundance ratios, each of the graphs has the same range in dex. 

O, Ne, S and Ar are $\alpha$-elements, which are synthesized in massive stars by $\alpha$-particle capture and released to the ISM by core-collapse supernovae. Cl is also produced by massive stars due to single proton or neutron captures by isotopes of  $\alpha$-elements. Therefore, the abundances of O, S, Ne, Ar and Cl should vary in lockstep, implying that abundance ratios of S, Ne, Ar and Cl with respect to O should be constant. We can see in Fig.~\ref{fig8} that these abundance ratios, S/O, Ne/O, Ar/O, and Cl/O, are very similar in both clouds, especially if N44C is excluded. N44C, which is the LMC \ion{H}{ii} region with the lowest value of O/H, is the only object with \ion{He}{ii} emission and the biases that can be introduced by the ICFs and by the assumed temperature structure can be different for this object. Excluding this object, the standard deviations of these relative abundances in both clouds based on the UVES/VLT data are very low, in the range 0.02--0.03~dex, similar to the dispersions in the oxygen abundances in each MC. The abundance ratios calculated with the previous spectra imply much larger standard deviations, in the range 0.07--0.15~dex. 

Considering the uncertainties, the S/O, Ne/O, and Ar/O ratios measured in the MCs are consistent with the solar values \citep{Lodd19}. The largest discrepancy is for Cl/O, but the difference between the two values is lower than two sigma.

\subsection{The N/O abundance ratio}\label{subsec5}

\begin{figure}
	\center
	\includegraphics[width=0.42\textwidth, trim=10 7 5 5, clip=yes]{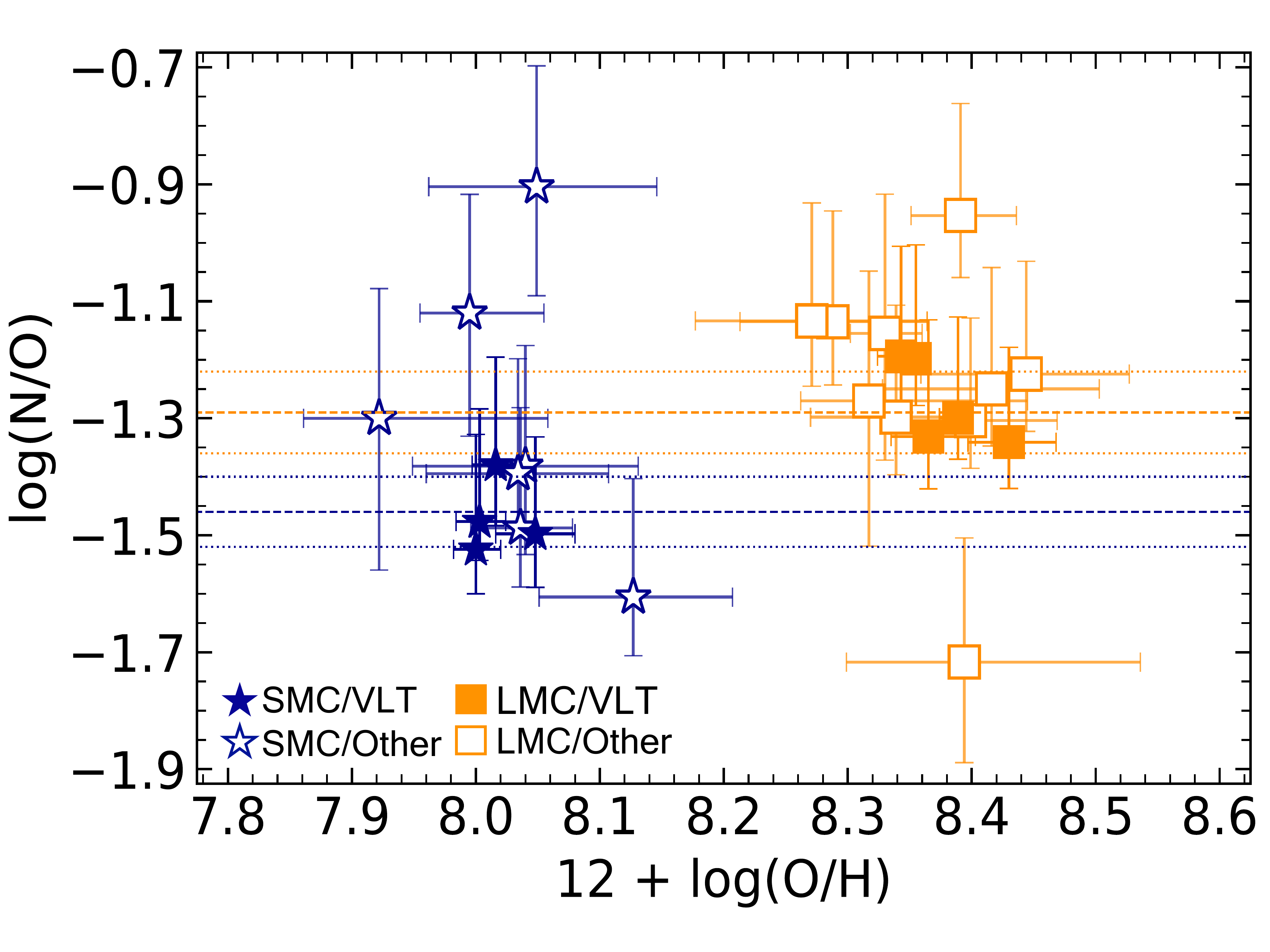}
	\caption{N/O abundance ratio as a function of O/H. Squares and stars are \ion{H}{ii} regions in the LMC and the SMC, respectively. Filled symbols show the results obtained with UVES/VLT spectra; empty symbols are for the results based on other spectra from the literature. The long and small dashed lines show the weighted means and standard deviations of the nitrogen abundances derived with UVES/VLT data in each cloud.}
	\label{fig9}
\end{figure}

The N/O abundance ratio as a function of the total oxygen abundance is presented in Fig.~\ref{fig9}. This abundance ratio is clearly different in the two MCs, $\log(\mbox{N}/\mbox{O})\simeq-1.25$ in the LMC and $\log(\mbox{N}/\mbox{O})\simeq-1.45$ in the SMC, presumably because of differences in the chemical evolution of each galaxy. The standard deviations obtained with the UVES/VLT spectra, which are equal to $0.05$~dex in both clouds, are fairly small in comparison with the ones implied by the previous spectra: 0.17~dex in the LMC and 0.20~dex in the SMC. Because of these large dispersions, previous analysis found similar values of N/O in both clouds \citep{PEFW78}. The N/O abundance ratios of the MCs are lower than the solar value \citep{Lodd19} by 0.42~dex and 0.59~dex for the SMC and LMC, respectively.

On the other hand, the standard deviations of 0.04--0.05~dex calculated with the UVES/VLT spectra for N/O and N/H are higher than the standard deviations of 0.02--0.03~dex obtained for O, S, Ne, Cl, and Ar. This is an expected behavior due to the complex nucleosynthetic origin of N, which can be produced both by massive stars and by stars of low and intermediate mass,\footnote{The relative contribution of both production channels is a controversial issue \citep[see, e.g.,][]{Prantzos11,Roy21}.} whose ejecta have very different mixing efficiencies \citep{Kru18,Eme18,Eme20}.

\subsection{The Fe/O abundance ratio}\label{subsec6}

The results for the iron abundances present a much larger range of variation from \ion{H}{ii} region to \ion{H}{ii} region than the abundances of the other elements. Besides, the [\ion{Fe}{iii}] lines on which the results are based are weaker than the other lines used in the abundance calculations, and the results provided by the deep UVES/VLT spectra are even better for iron than for the other elements. In fact, for IC~2111, NGC~1714, N81, and N90, we report for the first time the measurement of [\ion{Fe}{iii}] lines in the optical range. In N88A, we detect 16 [\ion{Fe}{iii}] lines, whereas \citet{Kurt99} only reported two. We can estimate the iron abundance in four objects using previous spectra, and the results differ from those derived with the UVES/VLT spectra by 0.02 to 0.24~dex.

\begin{figure}
	\center
	\includegraphics[width=0.42\textwidth, trim=10 7 5 5, clip=yes]{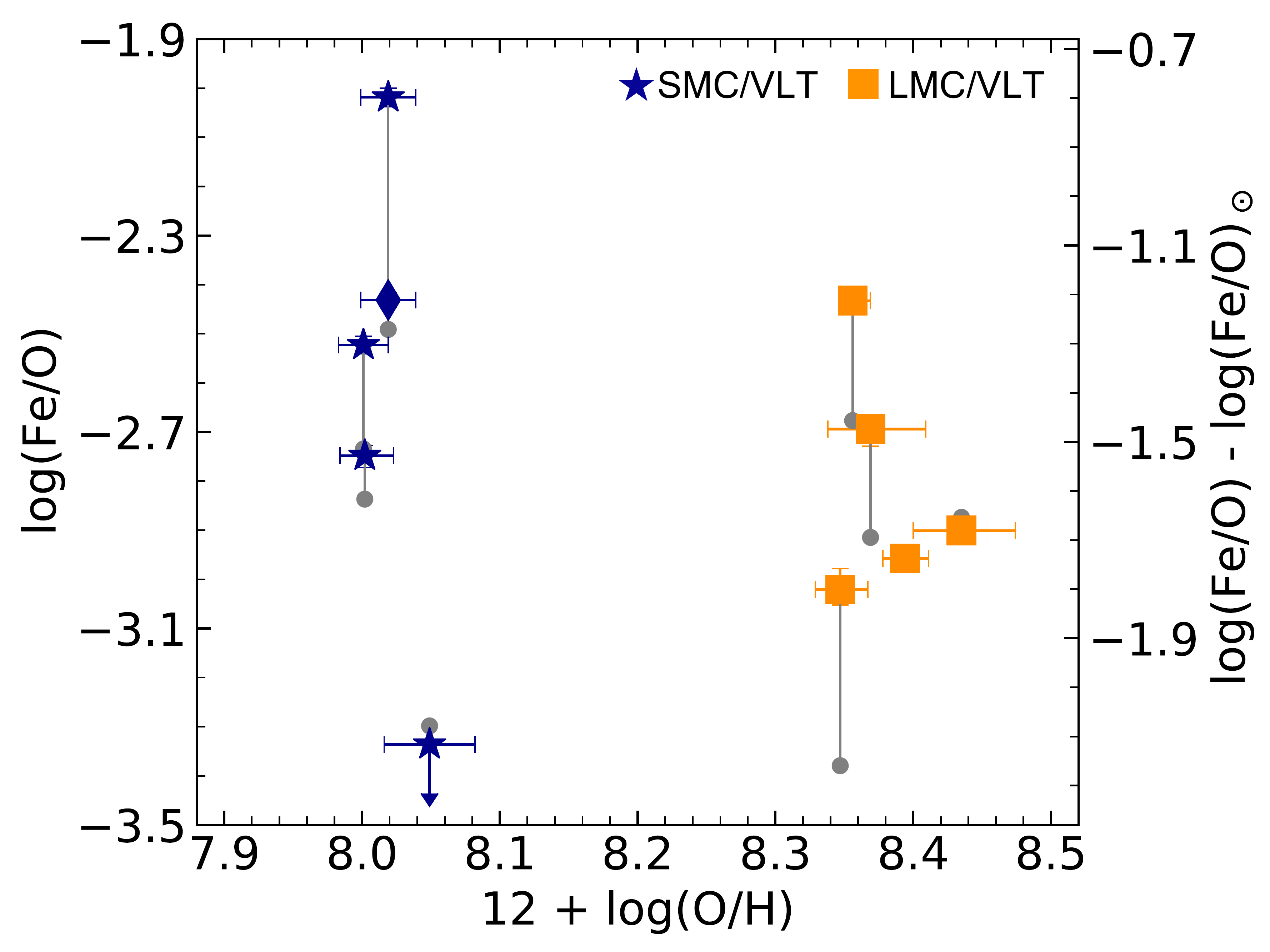}
	\caption{Fe/O abundance ratio (left axis) and the iron depletion factor (right axis) as a function of O/H. Squares and stars are regions in the LMC and the SMC and show results based on the ICF from equation~(2) of \citet{RR05}; the small circles use the ICF from their equation~(3). The diamond shows the iron abundance obtained by summing the ionic abundances of Fe$^{++}$ and Fe$^{3+}$ in N88A.}
	\label{fig10}
\end{figure}

Fig.~\ref{fig10} shows the values obtained with the UVES/VLT spectra for the Fe/O abundance ratio (left axis) as a function of O/H for all the objects in the sample. Squares and stars show the results for the LMC and the SMC, respectively. The squares and stars are connected to gray circles to show the values obtained with the two ICFs from equation~(2) and (3) of \citet{RR05}. In N88A we could measure two [\ion{Fe}{iv}] lines, and the total abundance can also be obtained by adding the ionic abundances of Fe$^{++}$ and Fe$^{3+}$. This procedure gives a value of $12 + \log(\mbox{Fe/H})=5.76\pm0.04$, which lies between the values obtained with the two ICFs and is shown in Fig.~\ref{fig10} with a diamond. 

The axis to the right of Fig.~\ref{fig10} shows the iron depletion factor calculated as $[\mbox{Fe/O}] = \log(\mbox{Fe/O})-\log(\mbox{Fe/O})_{\sun}$, with $\log(\mbox{Fe/O})_{\sun}=-1.28$ \citep{Lodd19}. The results presented in Fig.~\ref{fig10} show that the depletion factors of Fe/O in the LMC \ion{H}{ii} regions are similar to those found in Galactic nebulae \citep{DI+11}, with more than 90 per cent of their iron atoms condensed onto dust grains. The SMC \ion{H}{ii} regions have a similar behaviour, but with a wider range of iron depletions. These large variations in the iron abundance contrast with the small variations shown by the oxygen abundance in each MC. Note, however, that most of the iron atoms are located in dust grains whereas most of the oxygen atoms are in the gas phase. This means that the destruction of a small quantity of dust can change the gaseous iron abundance by a significant amount while changing the gaseous oxygen abundance by an unnoticeable amount.
 
\subsection{The chemical homogeneity in the MCs}\label{subsec7}

The fact that the derived standard deviations in O/H are similar to the uncertainties of the oxygen abundance determinations in each individual nebula, 0.02--0.04 dex (see Table~\ref{table5}), also indicates that the intrinsic standard deviations of O/H in each MC should be lower than 0.02--0.03~dex (see Table~\ref{table6}). Therefore, we conclude that oxygen is very well mixed in the gaseous phase of the interstellar medium (ISM) in each galaxy. This disagrees with the tentative conclusion of \citet{Roman21} that the interstellar gas in the LMC increases its metallicity by 0.8~dex from the west to the east side of the galaxy. Since the five LMC \ion{H}{ii} regions we are considering here (see Fig.~\ref{fig1}) are located in the west (IC~2111, N11B, and NGC~1714), middle (N44C), and east regions of the galaxy (30~Dor), the remarkable homogeneity in O/H shown by these regions suggests that the different behaviour of element abundances found by \citet{Roman21} is not due to metallicity variations but to differences in dust depletion, as the authors themselves propose.

Recent results for the Milky Way also reveal the high degree of homogeneity of the ISM in the Galactic disc \citep[see, e.g.,][]{Esteban22}. \citet{Arellano21} find a dispersion  of 0.07~dex around the radial gradient of O/H, only slightly larger than the typical uncertainties of the abundance determinations for the individual \ion{H}{ii} regions. Studies of the O/H gradients based on direct determinations of the electron temperature in the spectra of \ion{H}{ii} regions in M31, M33, M101 and other nearby spiral galaxies \citep[see, e.g., ][]{Bresolin11,Croxall16,TSC16,Esteban20,Rogers21,Rogers22} find dispersions around the O/H gradient fit between 0.04 and 0.10~dex, also consistent with the observational errors in the abundances. As in the present paper, the use of deep spectra that allow a good determination of the electron temperature and the O/H ratio is the common characteristic of those works. 

Analytic and semianalytic models and hydrodynamics simulations of the chemical enrichment in the interstellar medium of galaxies predict that dwarf galaxies should show larger dispersions in their chemical abundances than spiral galaxies \citep{Kru18,Eme18,Eme20}. Besides, \citet{Eme18} predict dispersions larger than one~dex when the metals are originated in the winds of asymptotic giant branch stars and dispersions of 0.5~dex when the elements come from supernovae. The dispersions we find in the MCs are much lower than these values, suggesting that the models might need to adjust their assumptions for parameters such as the fraction of metals produced that are retained in the disc or the gas velocity dispersions of their modelled galaxies.

\section{Conclusions}\label{sec7}

We present an analysis of deep echelle spectra taken with UVES at the VLT of eight \ion{H}{ii} regions in the MCs: IC~2111, N11B, N44C, and NGC~1714 in the LMC and N66A, N81, N88A, and N90 in the SMC. The spectra have a spectral resolution of 11600 (20000 for NGC~1714), and allow us to measure 92--225 lines in each object in the wavelength range 3100--10400~\AA.

The spectra are initially corrected for reddening using the extinction law derived by \citet{H83} for the LMC, with $R_{\mbox{v}}=3.1$. This law works well for all objects excepting N90 and N88A. In these two SMC regions, the law of \citet{OD94} with $R_{\mbox{v}}=5.5$ is used because it provides a much better fit to the relative intensities of the \ion{H}{i} lines.

We derive the physical conditions and ionic and total abundances of He, O, N, S, Ne, Ar, Cl, and Fe in all the regions. The same analysis is performed with one previous UVES/VLT spectra of the LMC \ion{H}{ii} region 30~Dor \citep{PA03} and for 18 previous spectra of the same MC objects.

With the UVES/VLT spectra we find average values of $12+\log(\mbox{O/H})=8.37$ in the LMC and $8.01$ in the SMC, and standard deviations of 0.02--0.03~dex. Similar standard deviations are found for the values of S/H, Ne/H, Ar/H, and Cl/H in each cloud and for the S/O, Ne/O, Ar/O, and Cl/O abundance ratios in both clouds. Because of the uncertainties in the derived abundances, the real dispersions can be expected to be lower. This result indicates that the chemical elements of the ISM in both, LMC and SMC are well mixed, at least at the level of the observational uncertainties.  

The N/O abundance ratio is $\sim0.20$~dex higher in the LMC than in the SMC. The standard deviations in the N/H and N/O abundance ratios in each cloud, which are equal to 0.04--0.05~dex, are also higher than those found for abundance ratios involving O, S, Ne, Ar, and Cl. Since nitrogen is mainly produced by asymptotic giant branch stars, this result is consistent with the predictions of models and simulations that find larger spreads for elements produced by less energetic events \citep{Kru18,Eme18,Eme20}.

The abundance ratios derived with the previous non-UVES/VLT spectra have average values similar to those found with the UVES/VLT spectra, except for Cl, but much larger dispersions. The standard deviations in the abundances of He, O, N, S, Ne, Ar, and Cl relative to H or O are equal to 0.04--0.20~dex for the results obtained with the non-UVES/VLT spectra, whereas the values based on UVES/VLT data are in the range 0.007--0.05~dex. This indicates the unprecedented quality of the data and chemical abundance results obtained in this work.

Finally, we find that the iron depletion factors are similar in the LMC and Galactic \ion{H}{ii} regions, with more than 90 per cent of the iron atoms condensed onto dust grains. The SMC \ion{H}{ii} regions behave in a similar way, but show more spread in their results.

\section*{Acknowledgements}
We thank the referee, R.\ B.\ C.\ Henry, for several useful suggestions. We also thank S.\ Sim\'on-D\'iaz for helpful comments on the results for the stellar abundances.This work is based on observations collected at the European Southern Observatory, Chile, with proposal numbers ESO 70.C-0008(A) and 92.C-0191(A). This research has made use of R band images from the `Aladin sky atlas', developed at CDS, Strasbourg Observatory, France. This research has used H$\alpha$ images from observations made with the NASA/ESA Hubble Space Telescope, obtained from the data archive at the Space Telescope Science Institute, which is operated by the Association of Universities for Research in Astronomy, Inc. under NASA contract NAS 5-26555. We acknowledge support from Mexican CONACYT grant CB-2014-240562. We acknowledge support from the Agencia Estatal de Investigaci\'on del Ministerio de Ciencia e Innovaci\'on (AEI-MCINN) under grant `Espectroscopia de campo integral de regiones \ion{H}{ii} locales. Modelos para el estudio de regiones \ion{H}{ii} extragal\'acticas' with reference 10.13039/501100011033. GDG acknowledges support from CONACYT grant 297932. JG-R acknowledges support from the Severo Ochoa excellence program SEV-2015-0548. JG-R and CE acknowledge support under grant P/308614 financed by funds transferred from the Spanish Ministry of Science, Innovation and Universities, charged to the General State Budgets and with funds transferred from the General Budgets of the Autonomous Community of the Canary Islands by the MCIU. LTSC acknowledges funding support from the Autonomous Community of Madrid through the project TEC2SPACE-CM (S2018/NMT-4291).


\section*{Data availability}
The data used in this work are available from the ESO archive facility at \texttt{http://archive.eso.org/}.
The line intensities measured in the spectra of all objects are available as supplementary material.

\bibliographystyle{mnras}
\bibliography{biblio}



\appendix

\section{Line intensities}\label{appendixA}
In this Appendix we present the line intensity ratios measured in eight \ion{H}{ii} regions of the Magellanic Clouds: IC~2111, N11B, N44C, NGC~1714, N66A, N81, N88A, and N90. For each emission line, we list the laboratory wavelength, $\lambda_0$, the emitting ion, the multiplet number (ID), the extinction law, $f(\lambda)$, the observed wavelength in the heliocentric framework, $\lambda$, and the line intensity ratio uncorrected, $F(\lambda)$, and corrected for extinction, $I(\lambda)$, with $F($H$\beta)=100$ and $I($H$\beta)=100$. For each region, we also list the observed intensity of H$\beta$ and the reddening coefficient $c(\mbox{H}\beta)$.
\include{table_lineints_LMC}
\include{table_lineints_SMC}

\section{Physical conditions, ionic and total abundances for the previously available spectra}\label{appendixB}

\begin{table*}
 \centering
 \caption{Physical conditions in units of cm$^{-3}$ ($n_\text{e}$) and K ($T_\text{e}$), ionic abundances in units of 12+log(X$^{+i}$/H$^+$) and total abundances in units of 12+log(X/H) of \ion{H}{ii} regions in the LMC. Results derived from data compiled from the literature. }
 \label{LMC_ext1}
 \begin{tabular}{l c c c c c c}
  \hline
  & 30~Dor$^{1}$ & 30~Dor$^{2}$ & 30~Dor$^{3}$ & IC~2111$^{2}$ & N11B$^{4}$ & N11B$^{5}$ \\
  \hline
  \multicolumn{7}{c}{Physical conditions}\\
  \hline
  $n_\text{e}$([\ion{O}{ii}]) & $460\pm30$ & -- & $420^{+280}_{-210}$ & -- & -- & -- \\
  $n_\text{e}$([\ion{S}{ii}]) & $420\pm30$ & $490^{+370}_{-240}$ & $400^{+340}_{-220}$ & $890^{+580}_{-370}$ & $950^{+2240}_{-630}$ & $170^{+250}_{-100}$ \\
  $n_\text{e}$([\ion{Cl}{iii}]) & $320^{+330}_{-190}$ & $3700^{+1700}_{-1300}$ & $560^{+1030}_{-320}$ &  -- & -- & -- \\
  $n_\text{e}$([\ion{Ar}{iv}]) & $300^{+1020}_{-90}$ & -- & $1400^{+1500}_{-800}$ & -- & -- & -- \\
  Adopted value & $430\pm140$  & $490^{+370}_{-240}$ & $470^{+250}_{-200}$ & $890^{+580}_{-370}$ & $950^{+2240}_{-630}$ & $170^{+250}_{-100}$ \\
  \multicolumn{7}{c}{}\\
\end{tabular}
\end{table*}

\begin{table*}
 \centering
 \caption{Physical conditions in units of cm$^{-3}$ ($n_\text{e}$) and K ($T_\text{e}$), ionic abundances in units of 12+log(X$^{+i}$/H$^+$) and total abundances in units of 12+log(X/H) of \ion{H}{ii} regions in the LMC.  Results derived from data compiled from the literature. }
 \label{sampleext}
 \begin{tabular}{l c c c c c c}
  \hline
  & N11B$^{1}$ & N44C$^{2}$ & N44C$^{3}$ & N44C$^{4}$ & N44C$^{5}$ & NGC~1714$^{6}$  \\
  \hline
  \multicolumn{7}{c}{Physical conditions}\\
  \hline
  $n_\text{e}$([\ion{O}{ii}]) & $130^{+210}_{-70}$ & -- & -- & -- & -- & -- \\
  $n_\text{e}$([\ion{S}{ii}]) & $80^{+230}_{-40}$ & $360^{+590}_{-220}$ & $20^{+80}_{-2}$ & $1470^{+790}_{-490}$ & $160^{+270}_{-90}$ & $270^{+280}_{-150}$ \\
  $n_\text{e}$([\ion{Cl}{iii}]) & $1980^{+1150}_{-890}$ & -- & -- & -- & $380^{+10500}_{-210}$  & -- \\
  $n_\text{e}$([\ion{Ar}{iv}]) & -- & -- & -- & -- & $330^{+1390}_{-90}$ & -- \\
  Adopted value & $140^{+110}_{-80}$ & $360^{+590}_{-220}$ & $20^{+80}_{-2}$ & $1470^{+790}_{-490}$ & $700^{+610}_{-360}$ & $270^{+280}_{-150}$ \\
  \multicolumn{7}{c}{}\\
\end{tabular}
\end{table*}

\begin{table*}
 \centering
 \caption{Physical conditions in units of cm$^{-3}$ ($n_\text{e}$) and K ($T_\text{e}$), ionic abundances in units of 12+log(X$^{+i}$/H$^+$) and total abundances in units of 12+log(X/H) of \ion{H}{ii} regions in the SMC.  Results derived from data compiled from the literature. }
 \label{sampleext}
 \begin{tabular}{l c c c c c c c}
  \hline
  & N66A$^{1}$ & N66A$^{2}$ & N66$^{3}$ & N81$^{4}$  & N88A sq. A$^{5}$ & N88A bar$^{5}$ & N88A$^{6}$   \\
  \hline
  \multicolumn{8}{c}{Physical conditions}\\
  \hline
  $n_\text{e}$([\ion{O}{ii}]) & -- & --& $50^{-240}_{-10}$ & -- & -- & -- & -- \\
  $n_\text{e}$([\ion{S}{ii}]) & $80^{+750}_{-4}$ & $90^{+270}_{-40}$ & $70^{+280}_{-20}$ & $380^{+350}_{-210}$ & $1380^{+2400}_{-900}$ & $10300^{+8600}_{-7500}$ & $2000^{+1330}_{-720}$  \\ 
  $n_\text{e}$([\ion{Cl}{iii}]) & -- & $4600^{+3700}_{-2200}$ & $4500^{+1900}_{-1500}$ & -- & -- & $2300^{+4900}_{-1400}$ & $2400^{+4900}_{-1400}$ \\
  Adopted value & $80^{+750}_{-4}$ & $80^{+270}_{-40}$ & $130^{+120}_{-80}$ & $380^{+350}_{-210}$ & $1380^{+2400}_{-900}$ & $2300^{+4900}_{-1400}$ & $2500^{+1700}_{-1200}$  \\
  \multicolumn{8}{c}{}\\
  $T_\text{e}$([\ion{O}{iii}])  & $12410^{+650}_{-610}$ & $12150\pm570$ & $12370^{+610}_{-560}$ & $12200^{+300}_{-280}$ & $13420\pm760$ & $14200^{+410}_{-490}$ & $14700^{+870}_{-850}$ \\
\end{tabular}
\end{table*}

\bsp	
\label{lastpage}
\end{document}

%% file: table_lineints_LMC.tex
\begin{table*}
 \caption{Observed and dereddened line intensity ratios for the LMC \ion{H}{ii} regions IC~2111 and N11B.} 
 \label{tabapen}
 \label{tableLMC1}
\begin{tabular}{c c c c c c c c c c}
 \hline
 &&&& \multicolumn{3}{c}{IC~2111}& \multicolumn{3}{c}{N11B} \\
 \cline{5-7} \cline{8-10}
 $\lambda_0$ (\AA)& Ion & ID &$f(\lambda)$&$\lambda$ (\AA)& $F(\lambda)$ & $I(\lambda)$ &$\lambda$ (\AA)& $F(\lambda)$ & $I(\lambda)$\\
 \hline 
 $3187.84 $ & \ion{He}{i} & 3 & $0.550$ & $3190.4$ & $2.47\pm0.17$ & $3.65\pm0.35$ & --- & --- & --- \\
 $3447.59 $ & \ion{He}{i} & 7 & $0.466$ & $3450.5$ & $0.194\pm0.058$ & $0.270\pm0.082$ & $3450.9$ & $0.244\pm0.014$ & $0.345\pm0.023$\\
 $3498.64 $ & \ion{He}{i} & 40 & $0.451$ & --- & --- & --- & $3502.0$ & $0.110\pm0.013$ & $0.153\pm0.018$\\
 $3512.52 $ & \ion{He}{i} & 38 & $0.447$ & --- & --- & --- & $3515.9$ & $0.128\pm0.014$ & $0.178\pm0.021$\\
 $3530.50 $ & \ion{He}{i} & 36 & $0.443$ & --- & --- & --- & $3533.9$ & $0.153\pm0.012$ & $0.213\pm0.019$\\
 $3554.42 $ & \ion{He}{i} & 34 & $0.436$ & $3557.4$ & $0.147\pm0.052$ & $0.200\pm0.072$ & $3557.8$ & $0.235\pm0.013$ & $0.324\pm0.021$\\
\hline
  \end{tabular}
\end{table*}

\newpage
\begin{table*}
 \contcaption{} 
 \label{tabapen}
 \begin{tabular}{c c c c c c c c c c}
 \hline 
 &&&& \multicolumn{3}{c}{IC~2111}& \multicolumn{3}{c}{N11B} \\
  \cline{5-7} \cline{8-10}
 $\lambda_0$ (\AA)& Ion & ID &$f(\lambda)$&$\lambda$ (\AA)& $F(\lambda)$ & $I(\lambda)$ &$\lambda$ (\AA)& $F(\lambda)$ & $I(\lambda)$\\
 \hline
 $4153.30 $ & \ion{O}{ii} & 19 & $0.208$ & $4156.8$ & $0.0435\pm0.0087$ & $0.050\pm0.010$ & $4157.2$ & $0.0444\pm0.0034$ & $0.0518\pm0.0040$\\
 $4168.97 $ & \ion{He}{i} & 52 & $0.202$ & $4172.5$ & $0.0379\pm0.0082$ & $0.044\pm0.010$ & $4173.0$ & $0.0504\pm0.0038$ & $0.0585\pm0.0045$\\
 $4243.97 $ & $[$\ion{Fe}{ii}$]$ & 21F & $0.177$ & --- & --- & --- & $4248.1$ & $0.0093\pm0.0014$ & $0.0103\pm0.0011$\\
 $4267.15 $ & \ion{C}{ii} & 6 & $0.169$ & $4270.8$ & $0.092\pm0.012$ & $0.104\pm0.013$ & $4271.2$ & $0.0824\pm0.0057$ & $0.0934\pm0.0066$\\
 $4287.40 $ & $[$\ion{Fe}{ii}$]$ & 7F & $0.163$ & $4291.0$ & $0.0476\pm0.0090$ & $0.053\pm0.010$ & $4291.5$ & $0.0238\pm0.0014$ & $0.0271\pm0.0012$\\
 $4303.82 $ & \ion{O}{ii} & 53 & $0.157$ & $4307.5$ & $0.0285\pm0.0073$ & $0.0319\pm0.0082$ & $4307.9$ & $0.0448\pm0.0034$ & $0.0503\pm0.0039$\\
\hline
 \end{tabular}
\end{table*}

\newpage
\begin{table*}
 \contcaption{} 
 \label{tabapen}
 \begin{tabular}{c c c c c c c c c c}
 \hline 
 &&&& \multicolumn{3}{c}{IC~2111}& \multicolumn{3}{c}{N11B} \\
 \cline{5-7} \cline{8-10}
 $\lambda_0$ (\AA)& Ion & ID &$f(\lambda)$&$\lambda$ (\AA)& $F(\lambda)$ & $I(\lambda)$ &$\lambda$ (\AA)& $F(\lambda)$ & $I(\lambda)$\\
 \hline
 $5412.00 $ & $[$\ion{Fe}{iii}$]$ & 1F & $-0.121$ & --- & --- & --- & $5417.4$ & $0.0146\pm0.0028$ & $0.0137\pm0.0027$\\
 $5517.71 $ & $[$\ion{Cl}{iii}$]$ & 1F & $-0.139$ & $5522.3$ & $0.463\pm0.030$ & $0.419\pm0.028$ & $5523.0$ & $0.4516\pm0.0084$ & $0.4078\pm0.0083$\\
 $5537.88 $ & $[$\ion{Cl}{iii}$]$ & 1F & $-0.143$ & $5542.5$ & $0.369\pm0.027$ & $0.333\pm0.025$ & $5543.1$ & $0.3359\pm0.0061$ & $0.3023\pm0.0063$\\
 $5599.77 $ & $[$\ion{Cr}{ii}$]$ &  & $-0.154$ & $5605.3$ & $0.043\pm0.013$ & $0.039\pm0.011$ & --- & --- & --- \\
 $5754.64 $ & $[$\ion{N}{ii}$]$ & 1F & $-0.180$ & $5759.4$ & $0.333\pm0.025$ & $0.293\pm0.023$ & $5760.1$ & $0.3118\pm0.0058$ & $0.2731\pm0.0064$\\
 $5868.00 $ & $[$\ion{Ni}{v}$]$? &  & $-0.200$ & $5872.9$ & $0.043\pm0.013$ & $0.037\pm0.011$ & --- & --- & --- \\
\hline
 \end{tabular}
\end{table*}


\newpage
\begin{table*}
 \contcaption{} 
 \label{tabapen}
 \begin{tabular}{c c c c c c c c c c}
 \hline 
 &&&& \multicolumn{3}{c}{IC~2111}& \multicolumn{3}{c}{N11B} \\
 \cline{5-7} \cline{8-10}
 $\lambda_0$ (\AA)& Ion & ID &$f(\lambda)$&$\lambda$ (\AA)& $F(\lambda)$ & $I(\lambda)$ &$\lambda$ (\AA)& $F(\lambda)$ & $I(\lambda)$\\
 \hline
 $8333.78 $ & \ion{H}{i} & P24 & $-0.515$ & $8340.8$ & $0.225\pm0.016$ & $0.156\pm0.014$ & $8341.6$ & $0.1850\pm0.0041$ & $0.1264\pm0.0055$\\
 $8359.00 $ & \ion{H}{i} & P22 & $-0.518$ & $8366.1$ & $0.305\pm0.018$ & $0.211\pm0.018$ & $8366.9$ & $0.2777\pm0.0060$ & $0.1897\pm0.0083$\\
 $8361.67 $ & \ion{He}{i} & 1/6 & $-0.518$ & $8368.8$ & $0.097\pm0.011$ & $0.0669\pm0.0088$ & --- & --- & --- \\
 $8374.48 $ & \ion{H}{i} & P21 & $-0.519$ & $8381.5$ & $0.377\pm0.021$ & $0.261\pm0.021$ & --- & --- & --- \\
 $8392.40 $ & \ion{H}{i} & P20 & $-0.521$ & --- & --- & --- & $8400.4$ & $0.3700\pm0.0081$ & $0.252\pm0.011$\\
 $8413.32 $ & \ion{H}{i} & P19 & $-0.523$ & $8420.4$ & $0.424\pm0.022$ & $0.292\pm0.023$ & $8421.4$ & $0.3798\pm0.0085$ & $0.258\pm0.011$\\
\hline
 \end{tabular}
\end{table*}

\begin{table*}
 \caption{Observed and dereddened line intensity ratios for the LMC \ion{H}{ii} regions N44C and NGC~1714.} 
 \label{tabapen}
 \label{tableLMC2}
 \begin{tabular}{c c c c c c c c c c}
 \hline
 &&&& \multicolumn{3}{c}{N44C}& \multicolumn{3}{c}{NGC~1714} \\
 \cline{5-7} \cline{8-10}
 $\lambda_0$ (\AA)& Ion & ID &$f(\lambda)$&$\lambda$ (\AA)& $F(\lambda)$ & $I(\lambda)$ &$\lambda$ (\AA)& $F(\lambda)$ & $I(\lambda)$\\
 \hline 
 $3132.79$ & \ion{O}{iii} & 3S-3P$_0$ & $0.571$ & $3136.0$ & $4.35\pm0.20$ & $5.66\pm0.35$ & ---  & --- & --- \\
 $3187.84$ & \ion{He}{i} & 3 & $0.550$ & $3190.9$ & $2.28\pm0.12$ & $2.93\pm0.19$ & $3190.9$ & $2.49\pm0.15$ & $3.79\pm0.34$ \\
 $3203.10$ & \ion{He}{ii} & 3.5 & $0.545$ & $3206.4$ & $4.75\pm0.21$ & $6.10\pm0.36$ & ---  & --- & --- \\
 $3299.39$ & \ion{O}{iii} & 3 & $0.511$ & $3302.7$ & $0.323\pm0.026$ & $0.408\pm0.036$ & ---  & --- & --- \\
 $3312.33$ & \ion{O}{iii} & 3 & $0.507$ & $3315.7$ & $0.615\pm0.042$ & $0.776\pm0.061$ & ---  & --- & --- \\
 $3340.77$ & \ion{O}{iii} & 3 & $0.498$ & $3344.2$ & $0.765\pm0.050$ & $0.962\pm0.072$ & ---  & --- & --- \\
\hline
  \end{tabular}
\end{table*}

\newpage
\begin{table*}
 \contcaption{} 
 \label{tabapen}
 \begin{tabular}{c c c c c c c c c c}
 \hline 
 &&&& \multicolumn{3}{c}{N44C}& \multicolumn{3}{c}{NGC~1714} \\
 \cline{5-7} \cline{8-10}
 $\lambda_0$ (\AA)& Ion & ID &$f(\lambda)$&$\lambda$ (\AA)& $F(\lambda)$ & $I(\lambda)$ &$\lambda$ (\AA)& $F(\lambda)$ & $I(\lambda)$\\
 \hline
 $3920.68$ & \ion{C}{ii} & 4 & $0.294$ & --- & --- & --- & $3924.7$ & $0.0328\pm0.0076$ & $0.041\pm0.010$ \\
 $3923.48$ & \ion{He}{ii} & 4.15 & $0.293$ & $3927.5$ & $0.0889\pm0.0070$ & $0.1018\pm0.0083$ & --- & --- & --- \\
 $3926.53$ & \ion{He}{i} & 58 & $0.292$ & $3930.5$ & $0.0861\pm0.0068$ & $0.0985\pm0.0080$ & $3930.6$ & $0.094\pm0.012$ & $0.117\pm0.015$ \\
 $3964.73$ & \ion{He}{i} & 5 & $0.277$ & $3968.7$ & $0.531\pm0.031$ & $0.604\pm0.038$ & $3968.8$ & $0.579\pm0.032$ & $0.715\pm0.046$ \\
 $3967.46$ & [\ion{Ne}{iii}]  & 1F & $0.276$ & $3971.5$ & $14.87\pm0.53$ & $16.89\pm0.69$ & $3971.5$ & $7.39\pm0.30$ & $9.12\pm0.47$ \\
 $3968.43$ & \ion{He}{ii} & 4.14 & $0.276$ & $3972.5$ & $0.1001\pm0.0077$ & $0.1137\pm0.0090$ & --- & --- & --- \\
\hline
 \end{tabular}
\end{table*}

\newpage
\begin{table*}
 \contcaption{} 
 \label{tabapen}
 \begin{tabular}{c c c c c c c c c c}
 \hline 
 &&&& \multicolumn{3}{c}{N44C}& \multicolumn{3}{c}{NGC~1714} \\
 \cline{5-7} \cline{8-10}
 $\lambda_0$ (\AA)& Ion & ID &$f(\lambda)$&$\lambda$ (\AA)& $F(\lambda)$ & $I(\lambda)$ &$\lambda$ (\AA)& $F(\lambda)$ & $I(\lambda)$\\
 \hline
 $4714.36$ & [\ion{Ne}{iv}] & 2D-2P & $0.038$ & $4719.0$ & $0.0569\pm0.0048$ & $0.0579\pm0.0049$ & --- & --- & --- \\
 $4724.15$ & [\ion{Ne}{iv}] & 1F & $0.035$ & $4729.0$ & $0.0486\pm0.0042$ & $0.0494\pm0.0043$ & --- & --- & --- \\
 $4725.62$ & [\ion{Ne}{iv}] & 1F & $0.035$ & $4730.4$ & $0.0372\pm0.0033$ & $0.0378\pm0.0034$ & --- & --- & --- \\
 $4733.93$ & [\ion{Fe}{iii}] & 3F & $0.032$ & --- & --- & --- & $4738.7$ & $0.0230\pm0.0066$ & $0.0236\pm0.0068$ \\
 $4740.16$ & [\ion{Ar}{iv}] & 1F & $0.031$ & $4745.0$ & $2.433\pm0.086$ & $2.468\pm0.087$ & $4745.1$ & $0.090\pm0.012$ & $0.092\pm0.012$ \\
 $4754.83$ & [\ion{Fe}{iii}] & 3F & $0.027$ & --- & --- & --- & $4759.6$ & $0.0530\pm0.0093$ & $0.0541\pm0.0095$ \\
\hline
 \end{tabular}
\end{table*}

\newpage
\begin{table*}
 \contcaption{} 
 \label{tabapen}
 \begin{tabular}{c c c c c c c c c c}
 \hline 
 &&&& \multicolumn{3}{c}{N44C}& \multicolumn{3}{c}{NGC~1714} \\
 \cline{5-7} \cline{8-10}
 $\lambda_0$ (\AA)& Ion & ID &$f(\lambda)$&$\lambda$ (\AA)& $F(\lambda)$ & $I(\lambda)$ &$\lambda$ (\AA)& $F(\lambda)$ & $I(\lambda)$\\
 \hline
 $7751.10$ & [\ion{Ar}{iii}] & 2F & $-0.457$ & $7758.9$ & $2.544\pm0.053$ & $2.061\pm0.081$ & $7759.1$ & $3.62\pm0.15$ & $2.55\pm0.17$ \\
 $7816.13$ & \ion{He}{i} & 1/7 & $-0.464$ & --- & --- & --- & $7824.2$ & $0.066\pm0.014$ & $0.047\pm0.010$ \\
 $8045.63$ & [\ion{Cl}{iv}] & 1F & $-0.488$ & $8053.8$ & $0.564\pm0.012$ & $0.451\pm0.019$ & $8054.1$ & $0.041\pm0.011$ & $0.0281\pm0.0080$ \\
 $8236.77$ & \ion{He}{ii} & 5.9 & $-0.506$ & $8245.2$ & $0.586\pm0.016$ & $0.464\pm0.021$ & --- & --- & --- \\
 $8245.64$ &\ion{H}{i} & P42 & $-0.507$ & --- & --- & --- & $8254.2$ & $0.044\pm0.012$ & $0.0302\pm0.0082$ \\
 $8247.73$ &\ion{H}{i} & P41 & $-0.507$ & --- & --- & --- & $8256.1$ & $0.050\pm0.012$ & $0.0338\pm0.0086$ \\
\hline
 \end{tabular}
\end{table*}

%% file: table_lineints_SMC.tex
\begin{table*}
 \caption{Observed and dereddened line intensity ratios for the SMC \ion{H}{ii} regions N66A and N81.} 
 \label{tabapen}
 \label{tableSMC1}
 \begin{tabular}{c c c c c c c c c c}
 \hline
 &&&& \multicolumn{3}{c}{N66A}& \multicolumn{3}{c}{N81} \\
 \cline{5-7} \cline{8-10}
 $\lambda_0$ (\AA)& Ion & ID &$f(\lambda)$&$\lambda$ (\AA)& $F(\lambda)$ & $I(\lambda)$ &$\lambda$ (\AA)& $F(\lambda)$ & $I(\lambda)$\\
 \hline 
 $3187.84$ & \ion{He}{i} & 3 & $0.550$ & $3189.48$ & $2.81\pm0.14$ & $3.68\pm0.26$ & $3189.4$ & $3.40\pm0.13$ & $3.81\pm0.26$ \\
 $3296.77$ & \ion{He}{i} & 9 & $0.512$ & --- & --- & --- & $3298.5$ & $0.0819\pm0.0045$ & $0.0911\pm0.0069$ \\
 $3342.50$ & [\ion{Ne}{iii}] & 1D-1S & $0.497$ & --- & --- & --- & $3344.3$ & $0.1615\pm0.0082$ & $0.179\pm0.013$ \\
 $3354.55$ & \ion{He}{i} & 8 & $0.494$ & --- & --- & --- & $3356.3$ & $0.1413\pm0.0073$ & $0.157\pm0.011$ \\
 $3447.59$ & \ion{He}{i} & 7 & $0.466$ & --- & --- & --- & $3449.4$ & $0.256\pm0.012$ & $0.282\pm0.019$ \\
 $3478.97$ & \ion{He}{i} & 43 & $0.457$ & --- & --- & --- & $3480.8$ & $0.0795\pm0.0043$ & $0.0875\pm0.0063$ \\
\hline
  \end{tabular}
\end{table*}


\newpage
\begin{table*}
 \contcaption{} 
 \label{tabapen}
 \begin{tabular}{c c c c c c c c c c}
 \hline 
 &&&& \multicolumn{3}{c}{N66A}& \multicolumn{3}{c}{N81} \\
 \cline{5-7} \cline{8-10}
 $\lambda_0$ (\AA)& Ion & ID &$f(\lambda)$&$\lambda$ (\AA)& $F(\lambda)$ & $I(\lambda)$ &$\lambda$ (\AA)& $F(\lambda)$ & $I(\lambda)$\\
 \hline
 $4026.21$ & \ion{He}{i} & 18 & $0.254$ & $4028.6$ & $1.98\pm0.12$ & $2.25\pm0.15$ & $4028.3$ & $1.881\pm0.076$ & $1.983\pm0.095$ \\
 $4068.60$ & [\ion{S}{ii}] & 1F & $0.238$ & $4071.0$ & $1.336\pm0.092$ & $1.50\pm0.11$ & $4070.8$ & $0.559\pm0.025$ & $0.587\pm0.030$ \\
 $4069.62$ & \ion{O}{ii} & 10 & $0.238$ & --- & --- & --- & $4071.9$ & $0.0506\pm0.0028$ & $0.0532\pm0.0032$ \\
 $4072.15$ & \ion{O}{ii} & 10 & $0.237$ & --- & --- & --- & $4074.3$ & $0.0198\pm0.0012$ & $0.0208\pm0.0014$ \\
 $4076.35$ & [\ion{S}{ii}] & 1F & $0.235$ & $4078.7$ & $0.445\pm0.040$ & $0.500\pm0.046$ & $4078.5$ & $0.200\pm0.010$ & $0.210\pm0.012$ \\
 $4101.74$ & \ion{H}{i} & H6 & $0.226$ & $4104.0$ & $22.31\pm0.79$ & $24.9\pm1.0$ & $4103.9$ & $24.21\pm0.86$ & $25.4\pm1.1$ \\
\hline
 \end{tabular}
\end{table*}


\newpage
\begin{table*}
 \contcaption{} 
 \label{tabapen}
 \begin{tabular}{c c c c c c c c c c}
 \hline 
 &&&& \multicolumn{3}{c}{N66A}& \multicolumn{3}{c}{N81} \\
 \cline{5-7} \cline{8-10}
 $\lambda_0$ (\AA)& Ion & ID &$f(\lambda)$&$\lambda$ (\AA)& $F(\lambda)$ & $I(\lambda)$ &$\lambda$ (\AA)& $F(\lambda)$ & $I(\lambda)$\\
 \hline
 $5978.93$ & \ion{Si}{ii} & 4 & $-0.218$ & --- & --- & --- & $5982.2$ & $0.0336\pm0.0014$ & $0.0321\pm0.0016$ \\
 $6046.44$ & \ion{O}{i} & 22 & $-0.229$ & --- & --- & --- & $6049.7$ & $0.0226\pm0.0011$ & $0.0215\pm0.0012$ \\
 $6300.30$ & [\ion{O}{i}] & 1F & $-0.269$ & $6304.1$ & $2.14\pm0.22$ & $1.87\pm0.20$ & $6303.8$ & $0.792\pm0.017$ & $0.749\pm0.026$ \\
 $6312.10$ & [\ion{S}{iii}] & 3F & $-0.271$ & $6315.5$ & $2.045\pm0.033$ & $1.791\pm0.054$ & $6315.4$ & $1.888\pm0.027$ & $1.784\pm0.055$ \\
 $6347.11$ & \ion{Si}{ii} & 2 & $-0.276$ & --- & --- & --- & $6350.6$ & $0.0494\pm0.0017$ & $0.0466\pm0.0021$ \\
 $6363.78$ & [\ion{O}{i}] & 1F & $-0.279$ & $6367.6$ & $0.69\pm0.10$ & $0.599\pm0.088$ & $6367.3$ & $0.2608\pm0.0069$ & $0.2460\pm0.0095$ \\
\hline
 \end{tabular}
\end{table*}

\newpage
\begin{table*}
 \contcaption{} 
 \label{tabapen}
 \begin{tabular}{c c c c c c c c c c}
 \hline 
 &&&& \multicolumn{3}{c}{N66A}& \multicolumn{3}{c}{N81} \\
 \cline{5-7} \cline{8-10}
 $\lambda_0$ (\AA)& Ion & ID &$f(\lambda)$&$\lambda$ (\AA)& $F(\lambda)$ & $I(\lambda)$ &$\lambda$ (\AA)& $F(\lambda)$ & $I(\lambda)$\\
 \hline
 $9068.90$ & [\ion{S}{iii}] & 1F & $-0.577$ & $9073.9$ & $18.03\pm0.55$ & $13.59\pm0.84$ & $9073.7$ & $15.99\pm0.49$ & $14.17\pm0.94$ \\
 $9123.30$ & [\ion{Cl}{ii}] & 1F & $-0.582$ & $9128.7$ & $0.0162\pm0.0016$ & $0.0122\pm0.0014$ & --- & --- & --- \\
 $9229.01$ & \ion{H}{i} & P9 & $-0.589$ & $9234.1$ & $2.935\pm0.090$ & $2.20\pm0.14$ & $9233.9$ & $2.558\pm0.078$ & $2.26\pm0.15$ \\
 $9530.60$ & [\ion{S}{iii}] & 1F & $-0.611$ & $9536.3$ & $31.70\pm0.97$ & $23.5\pm1.5$ & $9536.1$ & $27.90\pm0.85$ & $24.6\pm1.7$ \\
 $9545.97$ & \ion{H}{i} & P8 & $-0.612$ & $9551.2$ & $3.64\pm0.11$ & $2.70\pm0.18$ & $9551.0$ & $3.32\pm0.10$ & $2.92\pm0.20$ \\
 $9702.50$ & \ion{He}{i} & 3/7 & $-0.622$ & $9708.1$ & $0.0598\pm0.0044$ & $0.0441\pm0.0041$ & --- & --- & --- \\
 $9850.26$ & [\ion{C}{i}] & 1F & $-0.632$ & $9856.0$ & $1.045\pm0.050$ & $0.767\pm0.058$ & --- & --- & --- \\
 $10027.70$ & \ion{He}{i} & 6/7 & $-0.643$ & $10033.2$ & $0.290\pm0.016$ & $0.211\pm0.017$ & $10033.0$ & $0.221\pm0.012$ & $0.193\pm0.016$ \\
 $10336.40$ & [\ion{S}{ii}] & 3F & $-0.661$ & --- & --- & --- & $10341.9$ & $0.1437\pm0.0084$ & $0.125\pm0.011$ \\
 $10049.40$ & \ion{H}{i} & P7 & $-0.644$ & $10054.9$ & $7.80\pm0.24$ & $5.69\pm0.39$ & $10054.7$ & $6.86\pm0.21$ & $6.00\pm0.43$ \\
  \multicolumn{4}{l}{$c(\text{H}\beta)$}& \multicolumn{3}{c}{$0.21\pm0.04$}& \multicolumn{3}{c}{$0.09\pm0.04$} \\
  \multicolumn{4}{l}{$F(\text{H}\beta)$ (erg cm$^{-2}$ s$^{-1}$ $\mathring{\mbox{A}}^{-1}$)}& \multicolumn{3}{c}{$4.87\times10^{-13}$}& \multicolumn{3}{c}{$2.47\times10^{-12}$} \\
 \hline
 \end{tabular}
\end{table*}


\begin{table*}
 \caption{Observed and dereddened line intensity ratios for the SMC \ion{H}{ii} regions N88A and N90.} 
 \label{tabapen}
 \label{tableSMC2}
 \begin{tabular}{c c c c c c c c c c}
 \hline
 &&&& \multicolumn{3}{c}{N88A}& \multicolumn{3}{c}{N90} \\
 \cline{5-7} \cline{8-10}
 $\lambda_0$ (\AA)& Ion & ID &$f(\lambda)$&$\lambda$ (\AA)& $F(\lambda)$ & $I(\lambda)$ &$\lambda$ (\AA)& $F(\lambda)$ & $I(\lambda)$\\
 \hline 
$3109.18$ & $[$\ion{Ar}{iii}$]$ & 3P-1S & $0.176$ & $3110.7$ & $0.1897\pm0.0023$ & $0.2563\pm0.0067$ & --- & --- & --- \\
$3187.84$ & \ion{He}{i} & 3 & $0.181$ & $3189.3$ & $1.3297\pm0.0064$ & $1.812\pm0.044$ & --- & --- & --- \\
$3239.74$ & $[$\ion{Fe}{iii}$]$ & 6F & $0.183$ & $3241.4$ & $0.0524\pm0.0006$ & $0.0718\pm0.0019$ & --- & --- & --- \\
$3322.54$ & $[$\ion{Fe}{iii}$]$ & 5F & $0.185$ & $3324.1$ & $0.1154\pm0.0014$ & $0.1585\pm0.0043$ & --- & --- & --- \\
$3342.50$ & $[$\ion{Ne}{iii}$]$ & 1D-1S & $0.185$ & $3344.2$ & $0.2111\pm0.0025$ & $0.29\pm0.008$ & --- & --- & --- \\
$3354.55$ & \ion{He}{i} & 8 & $0.185$ & $3356.2$ & $0.0774\pm0.0009$ & $0.1063\pm0.0029$ & --- & --- & --- \\
\hline
 \end{tabular}
\end{table*}

\newpage
\begin{table*}
 \contcaption{} 
 \label{tabapen}
 \begin{tabular}{c c c c c c c c c c}
 \hline 
 &&&& \multicolumn{3}{c}{N88A}& \multicolumn{3}{c}{N90} \\
 \cline{5-7} \cline{8-10}
 $\lambda_0$ (\AA)& Ion & ID &$f(\lambda)$&$\lambda$ (\AA)& $F(\lambda)$ & $I(\lambda)$ &$\lambda$ (\AA)& $F(\lambda)$ & $I(\lambda)$\\
 \hline
 $4069.62$ & \ion{O}{ii} & 10 & $0.119$ & $4071.8$ & $0.0312\pm0.0010$ & $0.0382\pm0.0013$ & --- & --- & --- \\
$4072.15$ & \ion{O}{ii} & 10 & $0.118$ & $4074.2$ & $0.0303\pm0.0010$ & $0.0371\pm0.0013$ & --- & --- & --- \\
$4076.35$ & $[$\ion{S}{ii}$]$ & 1F & $0.118$ & $4078.4$ & $0.2223\pm0.0012$ & $0.2720\pm0.0045$ & $4078.7$ & $0.445\pm0.040$ & $0.482\pm0.044$ \\
$4101.74$ & \ion{H}{i} & H6 & $0.114$ & $4103.8$ & $21.20\pm0.75$ & $25.8\pm1.0$ & $4104.1$ & $23.77\pm0.86$ & $25.70\pm1.05$ \\
$4120.82$ & \ion{He}{i} & 16 & $0.112$ & $4122.9$ & $0.2551\pm0.0012$ & $0.3089\pm0.0048$ & $4123.1$ & $0.314\pm0.030$ & $0.339\pm0.033$ \\
$4143.76$ & \ion{He}{i} & 53 & $0.108$ & $4145.9$ & $0.2394\pm0.0012$ & $0.2884\pm0.0044$ & $4146.2$ & $0.246\pm0.025$ & $0.265\pm0.027$ \\
\hline
 \end{tabular}
\end{table*}


\newpage
\begin{table*}
 \contcaption{} 
 \label{tabapen}
 \begin{tabular}{c c c c c c c c c c}
 \hline
 &&&& \multicolumn{3}{c}{N88A}& \multicolumn{3}{c}{N90} \\
 \cline{5-7} \cline{8-10}
 $\lambda_0$ (\AA)& Ion & ID &$f(\lambda)$&$\lambda$ (\AA)& $F(\lambda)$ & $I(\lambda)$ &$\lambda$ (\AA)& $F(\lambda)$ & $I(\lambda)$\\
 \hline
 $5200.26$ & $[$\ion{N}{i}$]$ & 1F & $-0.052$ & $5202.9$ & $0.1035\pm0.0018$ & $0.0948\pm0.0018$ & --- & --- & --- \\
$5233.76$ & $[$\ion{Fe}{iv}$]$ & 4G-2F & $-0.056$ & $5236.2$ & $0.0257\pm0.0012$ & $0.0233\pm0.0011$ & --- & --- & --- \\
$5261.61$ & $[$\ion{Fe}{ii}$]$ & 19F & $-0.061$ & $5264.3$ & $0.0669\pm0.0013$ & $0.0603\pm0.0013$ & --- & --- & --- \\
$5270.40$ & $[$\ion{Fe}{iii}$]$ & 1F & $-0.062$ & $5273.2$ & $0.2407\pm0.0037$ & $0.2165\pm0.0038$ & --- & --- & --- \\
$5299.04$ & \ion{O}{i} & 26 & $-0.066$ & $5301.7$ & $0.0320\pm0.0012$ & $0.0286\pm0.0011$ & --- & --- & --- \\
$5333.65$ & $[$\ion{Fe}{ii}$]$ & 19F & $-0.071$ & $5336.4$ & $0.0166\pm0.0011$ & $0.0147\pm0.0010$ & --- & --- & --- \\
\hline
 \end{tabular}
\end{table*}

\newpage
\begin{table*}
 \contcaption{} 
 \label{tabapen}
 \begin{tabular}{c c c c c c c c c c}
 \hline
 & & & & \multicolumn{3}{c}{N88A}& \multicolumn{3}{c}{N90} \\
 \cline{5-7} \cline{8-10}
 $\lambda_0$ (\AA)& Ion & ID &$f(\lambda)$&$\lambda$ (\AA)& $F(\lambda)$ & $I(\lambda)$ &$\lambda$ (\AA)& $F(\lambda)$ & $I(\lambda)$\\
 \hline
 $8255.02$ & \ion{H}{i} & P38 & $-0.412$ & $8259.2$ & $0.0812\pm0.0008$ & $0.0401\pm0.0022$ & --- & --- & --- \\
$8257.85$ & \ion{H}{i} & P37 & $-0.412$ & $8262.1$ & $0.0824\pm0.0008$ & $0.0406\pm0.0022$ & --- & --- & --- \\
$8260.93$ & \ion{H}{i} & P36 & $-0.413$ & $8265.2$ & $0.0817\pm0.0008$ & $0.0403\pm0.0022$ & --- & --- & --- \\
$8264.28$ & \ion{H}{i} & P35 & $-0.413$ & $8268.6$ & $0.0848\pm0.0008$ & $0.0418\pm0.0023$ & --- & --- & --- \\
$8267.94$ & \ion{H}{i} & P34 & $-0.413$ & $8272.2$ & $0.0852\pm0.0008$ & $0.0420\pm0.0023$ & --- & --- & --- \\
$8286.43$ & \ion{H}{i} & P30 & $-0.415$ & $8290.7$ & $0.0773\pm0.0009$ & $0.0380\pm0.0021$ & --- & --- & --- \\
\hline
 \end{tabular}
\end{table*}